\documentclass[apj]{emulateapj}

\usepackage{threeparttable}
\usepackage{threeparttablex}
\usepackage{booktabs}
\usepackage{blindtext}
\usepackage{dashrule}

\usepackage{graphicx}
\usepackage{amssymb}
\usepackage{epstopdf}
\usepackage{float}
\usepackage[english]{babel}
\usepackage{amsmath}
\usepackage{booktabs}
\usepackage{textcomp}
\usepackage{verbatim}

\newcommand{\noop}[1]{}

\begin{document}

\title{The Origin of Molecular Clouds In Central Galaxies}

\author{F.A. Pulido\altaffilmark{1}, B.R. McNamara\altaffilmark{1,2}, A.C. Edge\altaffilmark{3}, M.T. Hogan\altaffilmark{1,2}, A.N. Vantyghem\altaffilmark{1}, H.R. Russell\altaffilmark{4}, P.E.J. Nulsen\altaffilmark{5,6}, I. Babyk\altaffilmark{1,7}, P. Salom\'e\altaffilmark{8}}
\affil{
$^{1}${Department of Physics and Astronomy, University of Waterloo, 200 University Ave W, Waterloo, ON N2L 3G1, Canada}\\
$^{2}${Perimeter Institute for Theoretical Physics, 31 Caroline St. North, Waterloo, ON N2L 2Y5, Canada} \\
$^{3}${Department of Physics, University of Durham, South Road, Durham DH1 3LE, United Kingdom} \\
$^{4}${Institute of Astronomy, Madingley Road, Cambridge CB3 0HA, UK} \\
$^{5}${Harvard-Smithsonian Center for Astrophysics, 60 Garden St., Cambridge, MA 02138, USA} \\
$^{6}${International Centre for Radio Astronomy Research (ICRAR), University of Western Australia, 35 Stirling Highway, Crawley, WA 6009, Australia} \\
$^{7}${Main Astronomical Observatory of NAS of Ukraine, 27 Akademika Zabolotnogo St, 03143, Kyiv, Ukraine} \\
$^{8}${Observatoire de Paris, LERMA, CNRS, PSL University, Sorbonne Univ. UPMC, Paris, France}
}

\begin{abstract}

We present an analysis of 55 central galaxies in clusters and groups with molecular gas masses and star formation rates lying between $10^{8}-10^{11}\ \text{M}_{\odot}$ and 0.5 and 270 $\text{M}_{\odot}\ \text{yr}^{-1}$, respectively. Using Chandra X-ray observations, we have calculated hydrostatic mass profiles, fully accounting for the central galaxy. We have derived acceleration profiles, atmospheric temperature, density, and other thermodynamic variables. Molecular gas mass is correlated with star formation rate,  H$\alpha$ line luminosity, and central atmospheric gas density. Molecular gas is detected only when the central cooling time or entropy index of the hot atmosphere falls below $\sim$1 Gyr or $\sim$35 keV cm$^2$, respectively, at a (resolved) radius of 10 kpc. These correlations indicate that the molecular gas condensed from hot atmospheres surrounding the central galaxies. The depletion timescale of molecular gas due to star formation approaches 1 Gyr in most systems. Yet ALMA images of roughly a half dozen systems drawn from this sample suggest the molecular gas formed recently and is in a transient state. We explore the origins of thermally unstable cooling by evaluating whether molecular gas becomes prevalent when the minimum of the cooling to free-fall time ratio ($\text{t}_{\text{cool}}/\text{t}_{\text{ff}}$) falls below $\sim10$. We find: 1) molecular gas-rich systems instead lie between $10 < \text{min}(\text{t}_{\text{cool}}/\text{t}_{\text{ff}}) < 25$, where $\text{t}_{\text{cool}}/\text{t}_{\text{ff}} = 25$ corresponds approximately to cooling time and entropy thresholds $\text{t}_{\text{cool}} \lesssim 1$ Gyr and $35~ \rm keV~cm^2$, respectively, 2) $\text{min}(\text{t}_{\text{cool}}/\text{t}_{\text{ff}}$) is uncorrelated with molecular gas mass and jet power, and 3) the narrow range $10 < \text{min}(\text{t}_{\text{cool}}/\text{t}_{\text{ff}}) < 25$ can be explained by an observational selection effect.  These results and the absence of isentropic cores in cluster atmospheres are in tension with ``precipitation" models, particularly those that assume thermal instability ensues from linear density perturbations in hot atmospheres.  Some and possibly all of the molecular gas may instead have condensed from atmospheric gas lifted outward either by buoyantly-rising X-ray bubbles or merger-induced gas motions.
\end{abstract}


\section{Introduction} \label{sec:intro}

Our understanding the origin and fate of molecular gas in galaxies is central to our understanding of galaxy formation. Large galaxy surveys such as the Sloan Digital Sky Survey firmly established the bimodality in the color distribution of local galaxies representing the so-called ``blue cloud" and ``red sequence" \citep{Baldry2004}. The blue cloud generally consists of star-forming spiral galaxies rich in molecular gas, while the red sequence is largely composed of quiescent elliptical galaxies \citep{Strateva2001}. This bimodality appears to result from an abrupt decrease in star formation that causes a rapid transition of galaxies from the blue cloud to the red sequence. \citep{Baldry2004,Faber2007,Thomas2005}. Star forming regions in galaxies tend to correlate with bright H$_2$ regions \citep{Leroy2008}.  Galaxies lacking star formation are also depleted in molecular gas. Thus, understanding the origin of molecular gas is critical to understanding of galaxy formation. 

Some of the largest reservoirs of molecular gas are found in central cluster galaxies, which are the most massive elliptical-like galaxies known. These galaxies, dubbed brightest cluster galaxies (BCGs), lie at the centres of galaxy clusters and groups.  Clusters and groups are embedded in hot, tenuous atmospheres whose temperatures lie between $10^{7-8}$ K. Their central cooling times are often less than the Hubble time.  The galaxies lying at their centers are expected to accumulate molecular gas that has condensed from the atmospheres \citep{Fabian1994}. Searches for this accumulated cool gas in clusters have covered a broad range of temperatures: soft X-ray emission \citep{Peterson2003}, ionized gas at $10^{5.5}$ K \citep{Bregman2006}, ionized gas at $10^4$ K \citep{Crawford1999}, neutral gas at $10^3$ K \citep{ODea1998}, warm molecular hydrogen gas at 1000-2500 K \citep{Edge2002}, and cold molecular hydrogen gas at 20-40 K \citep{Edge2001, Salome2003}.  Of these components, cold molecular hydrogen (which we refer to as molecular gas from here on), with masses lying between $10^{9-11}\ \text{M}_{\odot}$ far outweighs the others.  Early searches for molecular gas in clusters  resulted in H$_{2}$ upper limits of $\sim10^{8-10}\ \text{M}_{\odot}$ \citep{Bregman1988,Grabelsky1990,McNamara1994,ODea1994} and one detection in NGC1275 centered in the Perseus cluster \citep{Lazareff1989, Mirabel1989}. 

A breakthrough came with the detection of molecular gas in the central galaxies of twenty cooling clusters \citep{Edge2001, Salome2003} using IRAM 30m and JCMT 15m telescope observations of clusters selected from the ROSAT ALL-Sky Survey \citep{Crawford1999}. Although the molecular gas reservoirs are large, they account for less than $10\%$ of the mass expected from pure cooling models \citep{Edge2003}, indicating that cooling is suppressed.  Observations have since shown that radio-mechanical feedback from the active galactic nuclei (AGN) hosted by the BCG is the most plausible heating mechanism. In response to the cooling of the ICM, the AGN launches radio jets that inflate cavities and spawn shock waves in the surrounding atmospheres. The heat dissipated in the surrounding atmosphere via sound waves, shocks, turbulence, and cosmic rays nearly balance cooling  \citep[see reviews][]{McNamara2007,McNamara2012,Fabian2012}. Molecular gas is an essential element of feedback as it potentially links the fuel powering the AGN to the atmosphere that spawned it \citep{Gaspari2012,Pizzolato2005, McNamara2011}. 



In this paper, we investigate the origin of molecular gas in BCGs. Systems with nebular emission and recent star formation, both indirect tracers of molecular gas, are  preferentially found in cluster atmospheres  with short central cooling times, $\lesssim$ 1 Gyr, and low entropy indices, $\lesssim$ 30 keV cm$^{2}$, \citep{Cavagnolo2008,Rafferty2008}.  This cooling time threshold likely results from the onset of thermal instability at the centers of hot atmospheres \citep{Nulsen1986, Pizzolato2005}.  Some have suggested thermal instability ensues when the ratio of the cooling to free-fall time timescales $\text{t}_{\text{cool}}/\text{t}_{\text{ff}}$ falls below $\sim$10 \citep{Singh2015,McCourt2012,Gaspari2013,Prasad2015, Voit2015b, Li2015}.  However, \cite{McNamara2016} and \cite{Hogan2017} showed that this ratio is statistically governed almost entirely by the cooling time and not the free-fall time, casting doubt on this ratio as a useful thermodynamic parameter.  Furthermore they showed, as we do here, that  $\text{t}_{\text{cool}}/\text{t}_{\text{ff}}$ never falls significantly below 10 in BCGs, which is problematical for ``precipitation" models that assume thermal instability arises from linear density perturbations.  

New observations made with the  Atacama Large Millimeter Array (ALMA) have  resolved the molecular clouds in more that a half dozen systems \citep{McNamara2014,Russell2014,David2014,Russell2016b,Russell2016a,Vantyghem2016}.  In several systems, the molecular clouds lie in filamentary distributions located beneath the X-ray bubbles, indicating a direct link between molecular clouds and the AGN feedback process.  Unlike spiral galaxies, the molecular clouds in BCGs rarely lie in disks or rings.  These studies indicated that molecular gas is being lifted, or condensing, in the wakes of rising X-ray bubbles.  \cite{McNamara2016} proposed an alternative model where low entropy gas becomes thermally unstable when it is lifted to an altitude where its cooling time is much shorter than its infall time, $\text{t}_{\text{cool}}/\text{t}_{\text{I}} \lesssim 1$. Here the infall timescale, $t_{\rm I}$, is determined by the slower of the free-fall speed and the terminal speed of thermally unstable (see also \citet{Pizzolato2005}).  

Motivated by these considerations, we present an analysis of 55 giant elliptical galaxies situated in the cores of clusters and groups from which 33 are detected with  molecular gas. Section \ref{sec:data_sample} describes the sample consisting of systems observed with the IRAM 30m telescope. Section \ref{sec:analysis} describes the analyses taken to derive the molecular gas mass, ICM properties from \textit{Chandra} X-ray data, and cluster mass profiles following the procedure of \cite{Hogan2017b}. Section \ref{sec:results} and \ref{sec:discussion} present the results of these analyses and discussions regarding the connection of molecular gas with properties of the ICM and AGN.  

Throughout this paper, we have assumed a standard $\Lambda$CDM cosmology with $\Omega_{\rm m} = 0.3$, $\Omega_{\Lambda} = 0.7$, and H$_{0} = 70\ \text{km}\ \text{s}^{-1}\ \text{Mpc}^{-1}$. 
\section{Data Sample} \label{sec:data_sample}

\begin{figure}[t!]
\includegraphics[width=0.49\textwidth]{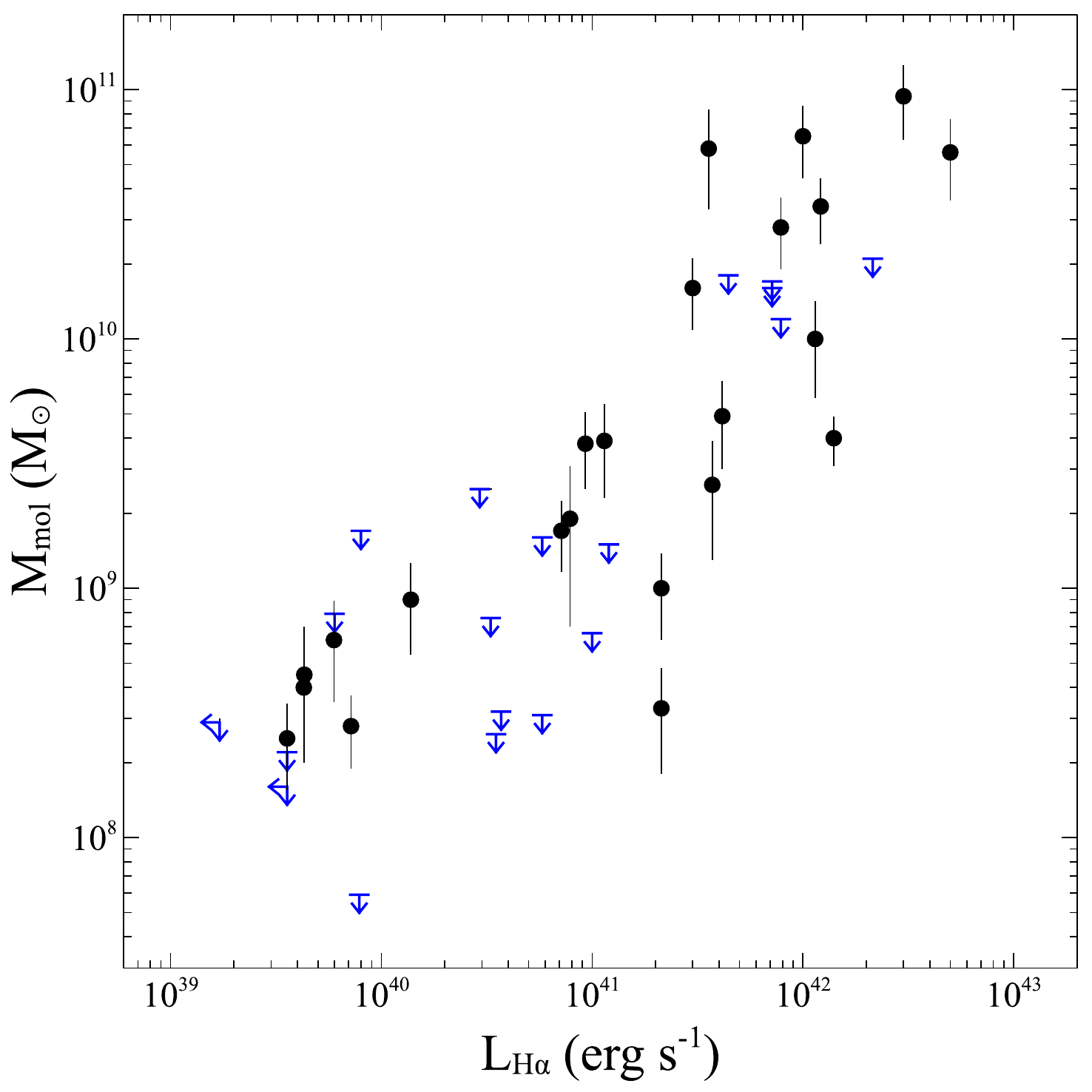}
\caption[Molecular gas mass vs. H$\alpha$ luminosity]{Molecular gas mass vs. H$\alpha$ luminosity for our sample. Black symbols denote systems observed with CO emission while blue symbols denote upper limits. \label{fig:molecular_gas_and_Halpha}}
\end{figure}

Our sample is composed of 55 central dominant galaxies drawn from the CO surveys of \cite{Edge2001}, \cite{Salome2003}, and others observed since these publications with the IRAM 30m by \cite{EdgePrep}. The sample was selected by a combination of properties including substantial mass cooling rate, and nebular emission \citep{Crawford1999}. The correlation between molecular gas mass and H$\alpha$ luminosity was previously found by \cite{Edge2001} and \cite{Salome2003}, and is apparent in our sample as shown in Figure \ref{fig:molecular_gas_and_Halpha}. We complement these earlier CO studies with X-ray data drawn from the \textit{Chandra} Data Archive by deriving mass profiles and other thermodynamic properties. Coordinates and X-ray observation properties for our sample are summarized in Tables \ref{tab:sample_bcg_coordinates} and \ref{tab:xray_observation_properties}, respectively. Our sample includes 33 systems detected in CO with derived molecular gas masses in the range $\sim$ $10^{8-11}\ \text{M}_{\odot}$, and 22 systems with CO upper limits.  Because our sample is not complete in a volume or flux limited sense, we avoid discussion of issues that may be affected by this bias.  Nevertheless, the sample represents the properties of systems over a four decade range of nebular luminosity and a three decade range of molecular gas mass (Figure \ref{fig:molecular_gas_and_Halpha}).

\vspace{1cm}
\section{Analysis} \label{sec:analysis}

We investigate the plausibility that molecular gas is condensing from hot atmospheres and is fueling star formation and AGN activity in central galaxies. We explore the properties of the surrounding hot atmospheres  and their relationship to the molecular gas observed in central galaxies. Section \ref{sub:molecular_gas_mass} discusses the molecular gas mass measurements and describes the analysis of Chandra X-ray analysis of the surrounding hot atmospheres. Cluster mass profiles were measured using Chandra X-ray and 2MASS infrared data as shown in Section \ref{sub:mass_profile}.

\subsection{Molecular Gas Mass}\label{sub:molecular_gas_mass}

All objects in our sample were observed with the IRAM 30m telescope. The cold molecular gas masses for several objects were taken from \cite{Edge2001} and \cite{Salome2003}, but corrected for a cosmology assuming H$_0$ = 70 km s$^{-1}$ Mpc$^{-1}$. Molecular gas masses for the remaining 26 objects were calculated using recent CO observations from \cite{EdgePrep}. CO detection for H1821+643 has also been made by \cite{Aravena2011} corresponding to a molecular gas mass of $\sim8.0\times10^{9}\ \text{M}_{\odot}$. We calcuated a molecular gas mass of $\sim 1\times10^{10}\ \text{M}_{\odot}$ using line intensity from \cite{EdgePrep}, and we use this value for our analysis. 

Line intensities taken from \cite{EdgePrep} were determined from measured antenna temperatures and velocity widths found from Gaussian fits to the CO spectra. For the IRAM 30m telescope, these were converted to integrated flux density $\text{S}_{\text{CO}}\Delta\nu$ using the following: 
\begin{equation}
   \text{S}_{\text{CO}}\Delta\nu(\text{Jy km s}^{\text{-1}}) = [6.8(1+z)^{-1/2}\ \text{Jy K}^{-1}]\text{I}_{\text{CO}}
\end{equation}

\noindent
where I$_{CO}$ is in units of K km s$^{-1}$ and $z$ is the redshift of the source. Integrated flux density in CO(2-1) or CO(3-2) was converted to an equivalent flux density in CO(1-2) assuming flux ratios CO(2-1)/CO(1-0) = 3.2 \citep{David2014} and CO(3-2)/CO(1-0) = 7.0 \citep{Russell2016a}. To translate integrated flux density in CO(1-0) directly to molecular gas mass we use the formulation taken from \citet{Bolatto2013}:

\begin{equation}\label{equation:molecular_gas_mass}
\footnotesize
  \text{M}_{\text{mol}} = 1.05 \times 10^{4} \left(\dfrac{\text{X}_{\text{CO}}}{2\times10^{20} \dfrac{\text{cm}^{-2}}{\text{K km s}^{-1}}}\right) \dfrac{\text{S}_{\text{CO}}\Delta\nu \text{D}_{\text{L}}^2}{(1+\text{z})}
\end{equation}

\vspace{4mm}
\noindent
where X$_{\text{CO}}$ is the CO-to-H$_{2}$ conversion factor, $\text{D}_{\text{L}}$ is the luminosity distance in Mpc. Molecular gas mass is sensitive to X$_{\text{CO}}$ which is not universal. We adopt the Galactic value X$_{\text{CO}} = 2 \times 10^{20}\ \text{cm}^{-2}\ (\text{K km s}^{-1})^{-1}$ with $\pm 30\%$ uncertainty following \cite{Bolatto2013} and previous studies of BCGs in cool core clusters \citep{Edge2001,Salome2003,Russell2014,McNamara2014,Russell2016a,Vantyghem2016}. This value is the mean conversion factor in the Milky Way galaxy. It is lower in the Galactic centre and higher at large radii. This X$_{\text{CO}}$ value can be approximately applied down to metallicities of $\sim$$0.5Z_{\odot}$ \citep{Bolatto2013}. The mean metallicity measured for the innermost regions of the objects in our sample is $0.66 \pm 0.38\ \text{Z}_{\odot}$. Therefore, adopting the Galactic value of X$_{\text{CO}}$ is reasonable.

\begin{deluxetable}{lccc}
  \tabletypesize{\scriptsize}
  \tablecaption{IRAM and ALMA Molecular Gas\label{tab:iram_and_alma}}
  \tablehead{
    \colhead{}        &             & IRAM                          & ALMA                        \\
    \colhead{System}  & Transition  & ($10^{8}$ M$_{\odot}$)        & ($10^{8}$ M$_{\odot}$)
  }
  \startdata
  2A0335+096          & CO(1-0)     & $17  \pm 5$     & $11.3 \pm 1.5$ \\
  A1664               & CO(1-0)     & $280 \pm 90$    & $110 \pm 10$   \\
  A1835               & CO(1-0)     & $650 \pm 210$   & $490 \pm 20$   \\
  A2597               & CO(1-0)     & $26 \pm 13$     & -              \\
                      & CO(2-1)     & $14 \pm 5$      & $18 \pm 2$     \\
  NGC5044             & CO(1-0)     & $2.3 \pm 0.8$   & -              \\
                      & CO(2-1)     & $0.61 \pm 0.2$  & $0.51$         \\
  PKS0745-191         & CO(1-0)     & $40 \pm 9$      & $46 \pm 9$    \\
  Phoenix             & CO(3-2)     & -               & $210 \pm 40$   \\
  \enddata
  \tablecomments{Molecular gas mass derived from IRAM observations shown above were calculated using data from \cite{EdgePrep}. References for ALMA observations: 2A0335+096-\cite{Vantyghem2016}, A1664-\cite{Russell2014} A1835-\cite{McNamara2014}, A2597-\cite{Tremblay2016}, NGC5044-\cite{David2014}, PKS0745-191-\cite{Russell2016a}, and Phoenix-\cite{Russell2016b}. An alternative name for 2A0335+096 is RXCJ0338.6+0958.}
\end{deluxetable}

Observations with the Atacama Large Millimeter Array (ALMA) have resolved the spatial and velocity structure of the molecular clouds in several objects in this sample. The molecular gas masses inferred from the IRAM and ALMA observations for the overlapping sample are compared in Table \ref{tab:iram_and_alma}. The molecular gas masses inferred from the CO(1-0) transition are generally larger for IRAM observations, suggesting the ALMA observations may have resolved away a fraction of extended CO emission.  However, the quality of the ALMA data is superior to the single dish IRAM data, and the measurements from the instruments are consistent to within their uncertainties. Nevertheless, it is reassuring that the two instruments are giving broadly consistent results. 



\subsection{X-ray Data Properties}\label{subsub:data_reduction}

The event data for all observations were obtained from the Chandra Data Archive (CDA). Each observation was reprocessed using the \textsc{chandra\_repro} script with \textsc{ciao} version 4.7. 

Events with bad grades were removed and background light curves were extracted from the level 2 event files. The events were filtered using the \textsc{lc\_clean} routine of M. Markevitch to identify and remove time intervals affected by flares. Blank-sky backgrounds were extracted using \textsc{caldb} version 4.6.7 for each observation, reprocessed identically to the event files, and normalized to match the 9.5-12.0 keV count rate in the observations. Calibrated event 2 files (and blank-sky backgrounds) were reprojected to match the position of the observation with the highest clean exposure time. Point sources were identified using \textsc{wavdetect} \citep{Freeman2002}, which were visually inspected and then excluded from further analysis.  

Spectra were extracted from concentric annuli forming spherical shells using \textsc{ciao} and binned to a minimum of 30 counts per channel. Following \cite{Hogan2017b}, we have taken the location of the BCG as the centre for our concentric annuli. To ensure that atmospheric temperature is measured accurately in deprojection, annuli were created with a minimum of $\sim$3000 net projected counts, with the number per annulus growing with increasing radius. Weighted redistribution matrix files (RMFs) and weighted auxiliary response files (ARFs) were created for each spectrum using the \textsc{mkacisrmf} and \textsc{mkwarf}, respectively. Lastly, the loss of area to chip gaps and point source extraction regions was corrected in the spectra. These spectra were then deprojected using a geometric routine \textsc{dsdeproj} described in \cite{Sanders2007} and \cite{Russell2008}.

\subsubsection{Spectral Fitting and Modelling the ICM}\label{subsub:fit}

Spectra were modeled with an absorbed single temperature \textsc{phabs(mekal)} thermal model \citep{Mewe1985,Mewe1986,Kaastra2015,Liedahl1995,Balucinska-Church1994} using \textsc{xspec} version 12.8.2 \citep{Arnaud1996}.  Abundances, anchored to the values in \cite{Anders1989}, were allowed to vary in the spectral fits. The hydrogen column density $\text{N}_{\text{H}}$ was frozen to the value of \cite{Kalberla2005} unless the best fit value was found to be significantly different. Fitting the spectra with the \textsc{phabs(mekal)} model yields values for temperature, metallicity, and \textsc{xspec} norm:

\begin{equation}\label{equation:norm}
	\text{norm} = \dfrac{10^{-14}}{4\pi \left(\text{D}_{\text{A}}(1+\text{z})\right)^{2}} \int \text{n}_{\text{e}}\text{n}_{\text{H}}\ \text{dV}
\end{equation} \\
\noindent
where z is redshift, D$_{\text{A}}$ is the angular distance to the source, n$_{\text{e}}$ and n$_{\text{H}}$ are the electron and hydrogen number densities, respectively. Augmenting the previous model to \textsc{phabs*cflux(mekal)}, we integrate the unabsorbed thermal model between $0.1-100$ keV and obtain an estimate for the bolometric flux of the X-ray emitting region. 

\subsubsection{Thermodynamic Properties of the Hot Atmosphere}\label{subsub:derived}

Electron density was computed using the normalization parameter of the thermal model. Assuming hydrogen and helium mass fractions of X = 0.75 and Y = 0.24, we find $n_{e} = 1.2 n_{H}$ \citep{Anders1989}. Taking $n_e$ and $n_H$ to be constant within each spherical shell, the electron density was computed from equation \ref{equation:norm}. Bolometric flux of the X-ray emitting region was converted to luminosity $\text{L}_{\text{X}}$. Pressure and entropy index were computed as $\text{P} = 2\text{n}_{\text{e}}\text{kT}$ and $\text{K} = \text{kTn}_{\rm e}^{-2/3}$, respectively. In a spherical shell of volume V, the cooling time of the ICM was computed as

\begin{equation}
\begin{split}
  \text{t}_{\text{cool}} &= \dfrac{3}{2}\dfrac{\text{P}}{\text{n}_{\text{e}}\text{n}_{\text{H}}\Lambda(\text{Z},\text{T})} = \dfrac{3}{2}\dfrac{\text{PV}}{\text{L}_{\text{X}}}
\end{split}
\end{equation}

\noindent
where $\Lambda(\text{Z},\text{T})$ is the cooling function as a function of metallicity Z and temperature T.  In a shell with electron density $\text{n}_{\text{e}}$, gas density was computed using $\rho_{\text{g}} = 1.2\text{n}_{\text{e}}\text{m}_{\text{p}}$ where $\text{m}_{\text{p}}$ is the mass of a proton. Finally, to obtain a radial gas mass distribution, gas density profiles were integrated in a piecewise manner from the centre of the cluster.

\subsection{Mass Profiles}\label{sub:mass_profile}
Hydrostatic mass profiles were derived and used to determine gravitational free-fall times and total cluster mass.  We adopted the mass model presented in \cite{Hogan2017b}. The model is composed of an NFW potential and a central cored isothermal potential representing the central galaxy,
\begin{equation}\label{equation:isonfwmass_potential}
\begin{split}
  \Phi_{\text{NFW}} &= -4 \pi \text{G} \rho_{0} \text{r}_{\text{s}}^{2} \dfrac{\ln{\left(1 + \text{r}/\text{r}_{\text{s}}\right)}}{\text{r}/\text{r}_{\text{s}}} \\
  \Phi_{\text{ISO}} &= \sigma^{2}\ln{\left( 1 + (\text{r}/\text{r}_{\text{I}})^{2} \right)},
\end{split}
\end{equation}  \\
\noindent
where $\rho_{0}$ is the characteristic gas density, r$_\text{s}$ is the scale radius of the NFW component, $\sigma$ is the stellar velocity dispersion, and r$_\text{I}$ is the scale radius of the isothermal component. The NFW profile \citep{Navarro1997} has been found to capture the total gravitating mass of clusters on large scales reasonably well \cite[e.g][]{Pointecouteau2005, Vikhlinin2006, Schmidt2007, Gitti2007, Babyk2014, Main2015}. However, the NFW profile alone underestimates the inferred mass from the observed velocity dispersion of stars in cluster cores \citep{Fisher1995, Lauer2014}. The gravitational potential is dominated by the stellar component in the innermost 10-20 kpc \citep{Li2012}. The isothermal component of this model accounts for the stellar mass of the central galaxy. 

This combined NFW and cored isothermal potential, dubbed \textsc{isonfwmass}, is implemented as an extension in the \textsc{xspec} package \textsc{clmass} \citep{Nulsen2010}. X-ray spectra derived from \textit{Chandra} data are fitted with this model, which assumes that the cluster is spherically symmetric and the gas is in hydrostatic equilibrium. To obtain a stable fit, \cite{Hogan2017b} set the r$_\text{I}$ parameter to an arbitrarily small but non-zero value and the $\sigma$ parameter frozen to an inferred stellar velocity dispersion $\sigma_{*}$ derived from 2MASS isophotal K-band magnitudes $\text{m}_{\text{k}20}$ measured within the isophotal radius $\text{r}_{\text{k20}}$. 

To determine the uncertainties of these quantities, we have utilized the \textsc{chain} command in \textsc{xspec} to generate a chain of sets of parameters via a Markov Chain Monte Carlo (MCMC) method. A chain length of 5000 was produced from which we adopted the standard deviation as the uncertainty of $\rho_{0}$, $\text{r}_{\text{s}}$, and mass profiles. Table \ref{tab:mass_properties} shows the fitted parameters for our mass profiles. The free-fall times $\text{t}_{\text{ff}}$ and total cluster mass proxy $\text{M}_{\Delta}$ were then computed as follows:
\begin{equation}\label{equation:freefall}
  \text{t}_{\text{ff}} = \sqrt{\dfrac{2\text{r}}{\text{g}}} \\
\end{equation}

\begin{equation}\label{equation:mdelta}
  \text{M}_{\Delta} = \dfrac{4\pi \text{R}_{\Delta}^{3}}{3} \Delta \rho_{\text{c}}
\end{equation}
\noindent
where g is the acceleration due to gravity and $\Delta=2500$. The values of M$_{2500}$ and R$_{2500}$ were determined from the combined NFW and isothermal profiles by numerically solving equation \ref{equation:mdelta}.

\begin{figure*}[ht!]
\centering
\includegraphics[width=0.49\textwidth]{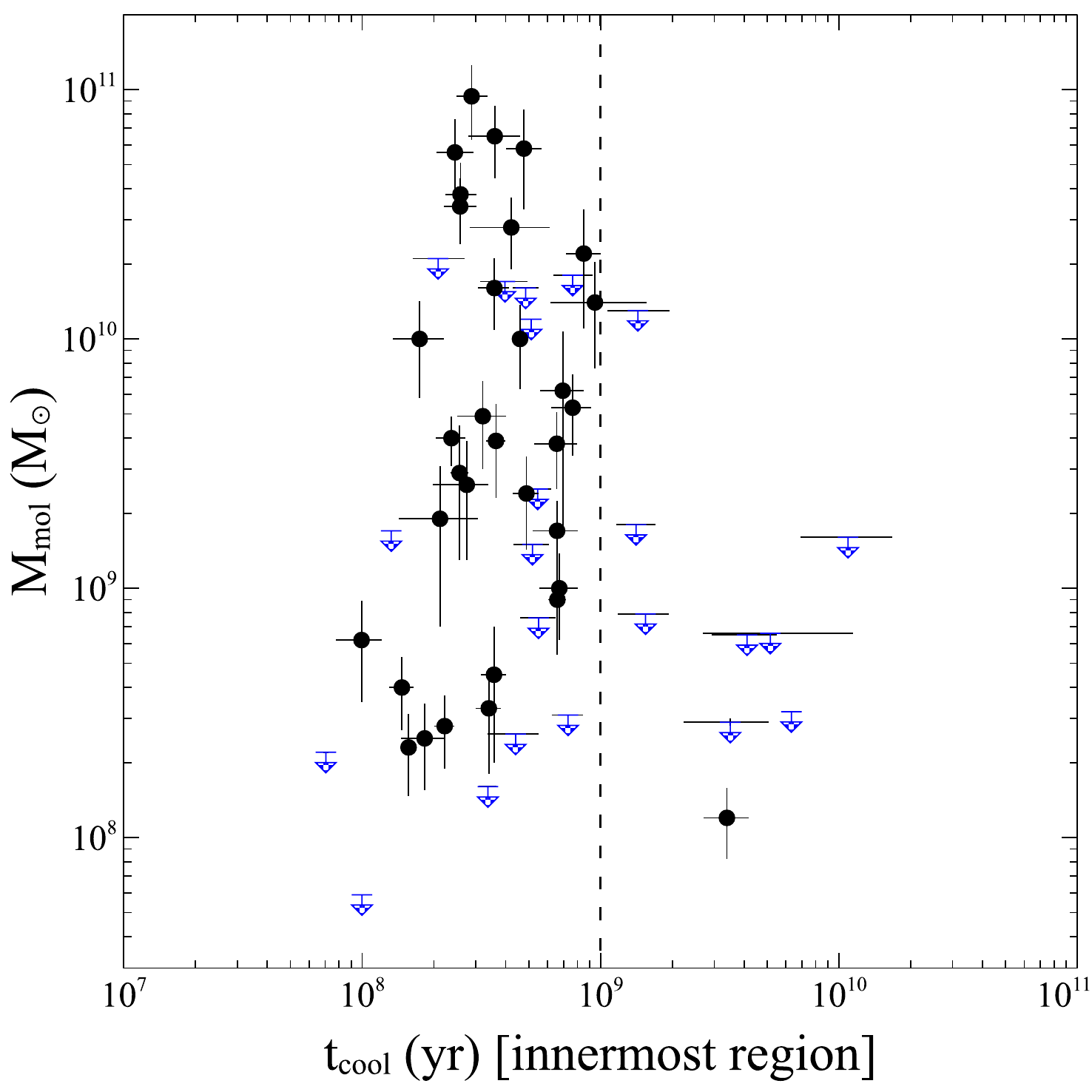}
\includegraphics[width=0.49\textwidth]{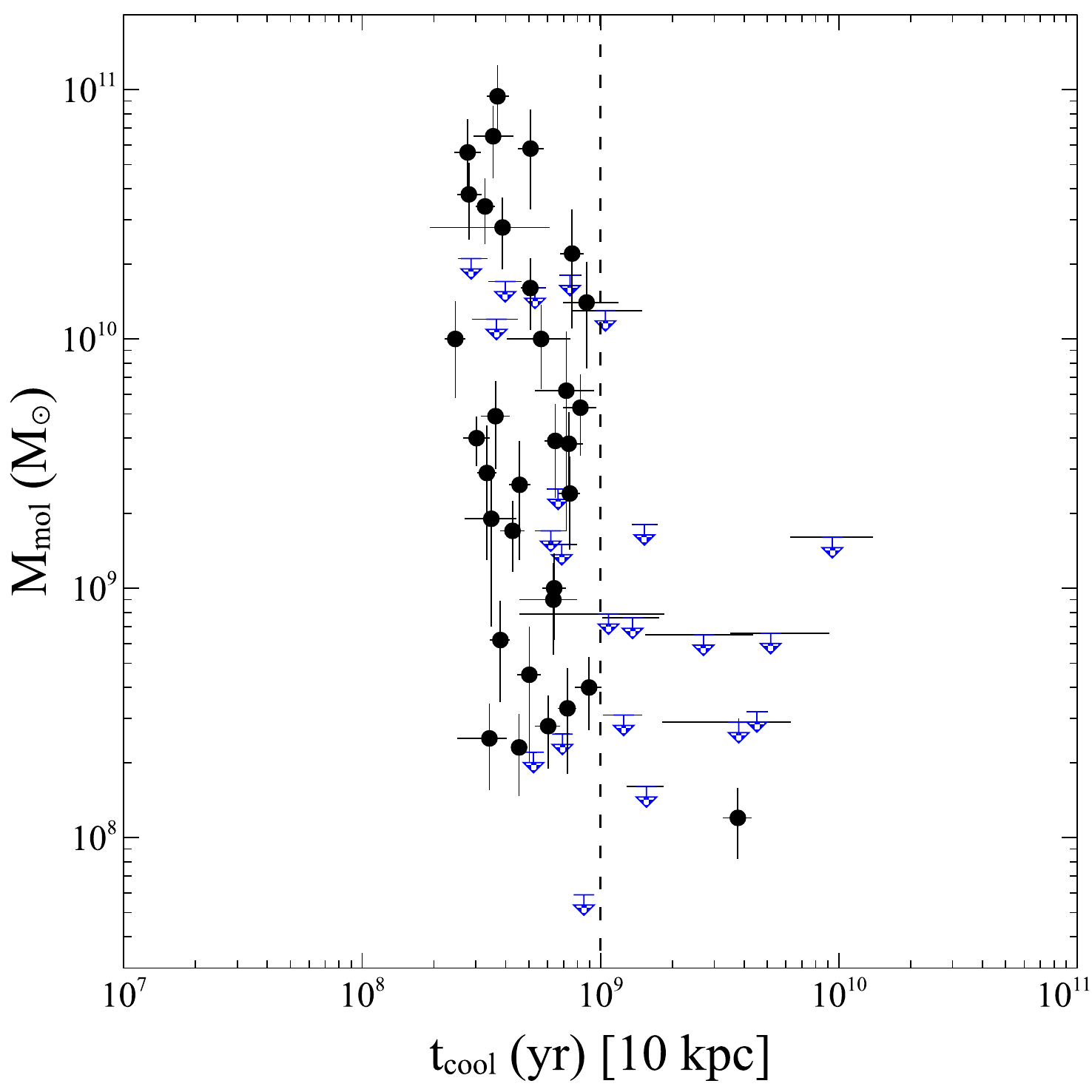}
\caption{Molecular gas mass vs. cooling time measured at the mean radius of the innermost region (left panel) and the cooling time measured at 10 kpc (right panel). Black symbols denote systems observed with CO emission while blue symbols denote upper limits. \label{fig:cooling_time_threshold}}
\end{figure*}

\subsection{AGN Mechanical Power}\label{sub:cavity_power}

Cavity powers were obtained from the literature for as many objects in our sample as possible (see Table \ref{tab:cavity_power_and_sfr} for references). Cavities inflated by radio-emitting jets from the central AGN allow a direct measurement of the mechanical energy output of the AGN. Assuming the cavities are in pressure balance with the surrounding atmosphere, the mean jet power required to create a cavity filled with relativistic gas is at least
\begin{equation}
  \rm P_{\rm cav} = \dfrac{\rm 4PV}{\rm t_{\rm age}}
\end{equation}
\noindent
where V is the volume of the cavity, P is the surrounding pressure, and $\rm t_{\rm age}$ is the age of the cavity, estimated by using the cavity's buoyancy time (time required for the cavity to rise buoyantly at its terminal velocity). In total we found cavity power measurements from literature for 27 objects in our sample.

A less reliable method for probing the mechanical output of an AGN is to use a correlation between its radio luminosity and cavity power. We derive mechanical power inferred from the AGN's radio luminosity using \citep{Birzan2008}
\begin{equation}
  \log{\text{P}_{\text{mech}}} = (0.48 \pm 0.07) \log{\text{L}_{\text{radio}}} + (2.32 \pm 0.09)
\end{equation}
\noindent
where the total radio luminosity was calculated by integrating the flux between $\nu_1 = 10$ MHz  and $\nu_2 = 5000$ MHz as
\begin{equation}
\begin{split}
  \text{L}_{\text{rad}} = 4\pi \text{D}_{\text{L}}^{2} \text{S}_{\nu_0} \int_{\nu_1}^{\nu_2} (\nu/\nu_{0})^{-\alpha} \text{d}\nu \\
\end{split}
\end{equation}
\noindent
following \cite{Birzan2004}. We used a spectral index of $\alpha \approx 0.75$ assuming a power-law spectrum S$_{\nu} \sim \nu^{-\alpha}$. We have taken the $\nu_{0} = $1400 MHz flux reported in the NRAO VLA Sky Survey (NVSS) Catalog \citep{Condon2002}. Table \ref{tab:cavity_power_and_sfr} shows the mechanical power inferred from radio luminosities for objects in our sample. \\

\section{Results} \label{sec:results}

\subsection{Cooling Time and Molecular Gas}\label{sub:cooling_time_molecular_gas}

In this section, we investigate whether and how the cooling time of the atmosphere is related to the molecular gas found in the central galaxies. Star formation and nebular emission in central galaxies ensues when the cooling time falls below $\sim 10^9$ yr, a phenomenon known as the cooling time threshold \citep{Rafferty2008, Cavagnolo2008, Hogan2017}. As star formation is strongly coupled to molecular gas mass, we study the relationship between molecular gas and cooling time.

Molecular gas mass is plotted against central cooling time in the left panel of Figure \ref{fig:cooling_time_threshold}. Because we have more than one molecular gas mass estimate for some of the systems in our sample and that we favour the most recent CO data from \cite{EdgePrep}, we plot the estimates calculated from CO line intensities taken from \cite{EdgePrep} when available. Otherwise, we plot the recomputed estimates from \cite{Edge2001} and \cite{Salome2003}.

\begin{figure}[b]
\centering
\includegraphics[width=0.49\textwidth]{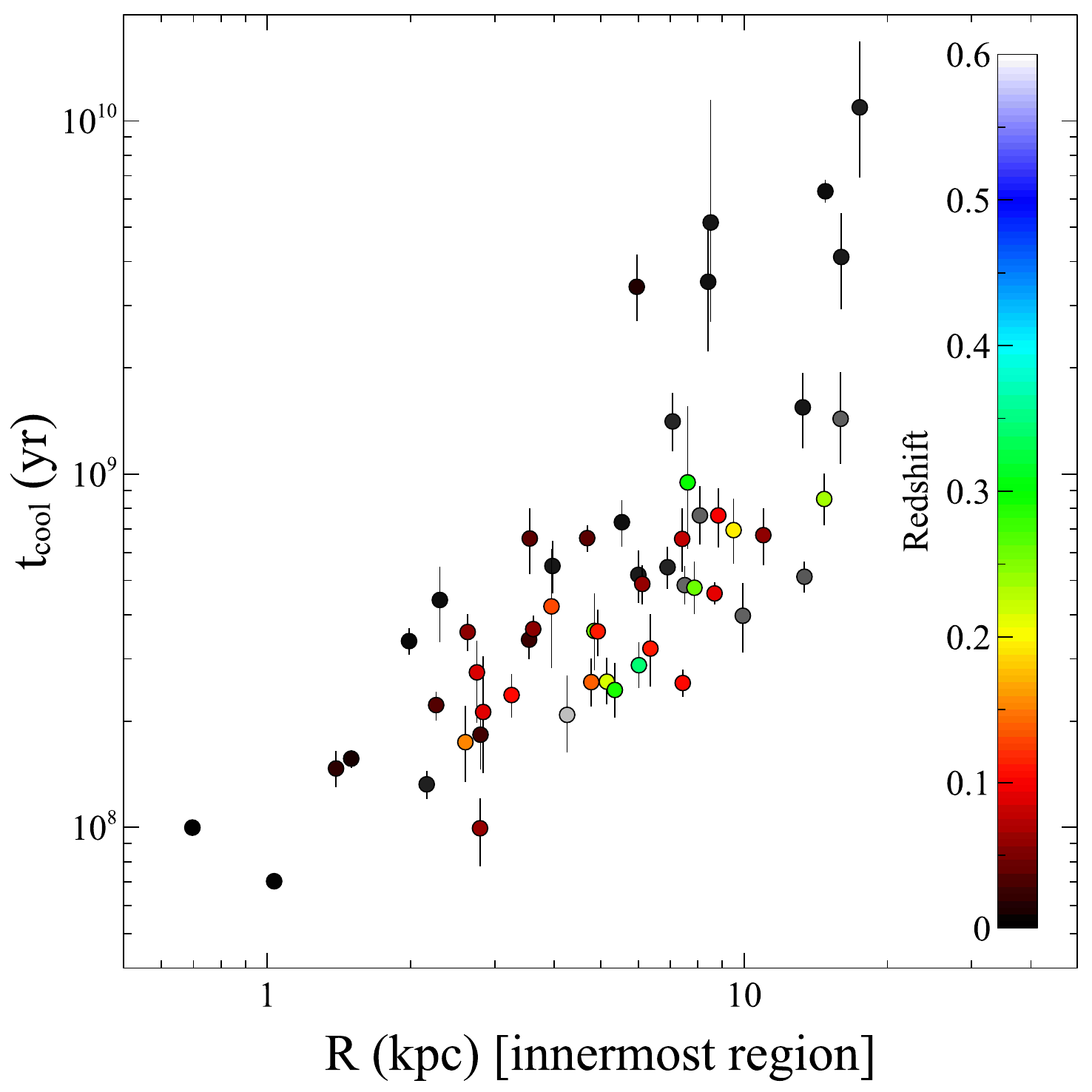}
\caption{Cooling time in the innermost region plotted against the mean radius of the innermost region. The tendency of the cooling time to increase with size of the innermost region is easily observed in this figure.\label{fig:resolution_effect_to_cooling_time}}
\end{figure}

Due to the large range in cluster distances the innermost bins do not sample the same linear diameters. This resolution effect is seen in Figure \ref{fig:resolution_effect_to_cooling_time}, which shows the central cooling time against the mean radius of the innermost region, $\text{R}_{\text{mid}} = (\text{R}_{\text{inner}} + \text{R}_{\text{outer}})/2$. The plot shows a tendency to measure lower central cooling times with smaller $\text{R}_{\text{mid}}$. This is consistent with the findings of \cite{Peres1998} and \cite{Hogan2017} who observed a trend of reduced central cooling time with increased resolution for the sample as a whole. To account for this resolution bias the right panel of Figure \ref{fig:cooling_time_threshold} shows the cooling time at a single physical radius of 10 kpc. Those systems whose innermost regions have $\text{R}_{\text{mid}} > 10$ kpc were extrapolated to 10 kpc using the linear slope of the first two points in the radial profile in log-log space. There are 8 such systems in our sample. Among these systems, the largest value found for $\text{R}_{\text{mid}}$ is 17.5 kpc.   When cooling times are plotted at a standard altitude of 10 kpc in the right panel of Figure \ref{fig:cooling_time_threshold}, the distribution of cooling times narrows, so that cooling times falling below $2 \times 10^{8}$ yr are not observed. This shows that the larger variation in cooling time seen in the left panel is due to resolution.

The sudden jump in molecular gas mass seen in Figure \ref{fig:cooling_time_threshold} shows that molecular gas resides preferentially in systems with central cooling times lying below $\sim$1 Gyr. Central galaxies located in clusters with longer central cooling times contain less than $\sim 10^9 ~M_\odot$ of molecular gas.  The molecular gas masses in central galaxies with atmospheric cooling times lying below 1 Gyr at 10 kpc rise dramatically to several $10^{10} ~M_\odot$.  Molecular gas masses of this magnitude dramatically exceed those in gas-rich spirals like the Milky Way.  Furthermore, we find a narrow spread in cooling time below $\sim$1 Gyr with a mean of 0.5 Gyr and a standard deviation of 0.2 Gyr. 

Figure \ref{fig:cooling_time_threshold} shows the same $\sim 1$ Gyr cooling time threshold for the onset of molecular gas that has previously been found in H$\alpha$ emission and star formation.  This threshold suggests that 
molecular gas is linked to hot atmospheres with short cooling times, consistent with the hypothesis that molecular clouds in central galaxies condense from hot atmospheres.  

\begin{figure*}[ht!]
\centering
\includegraphics[width=0.49\textwidth]{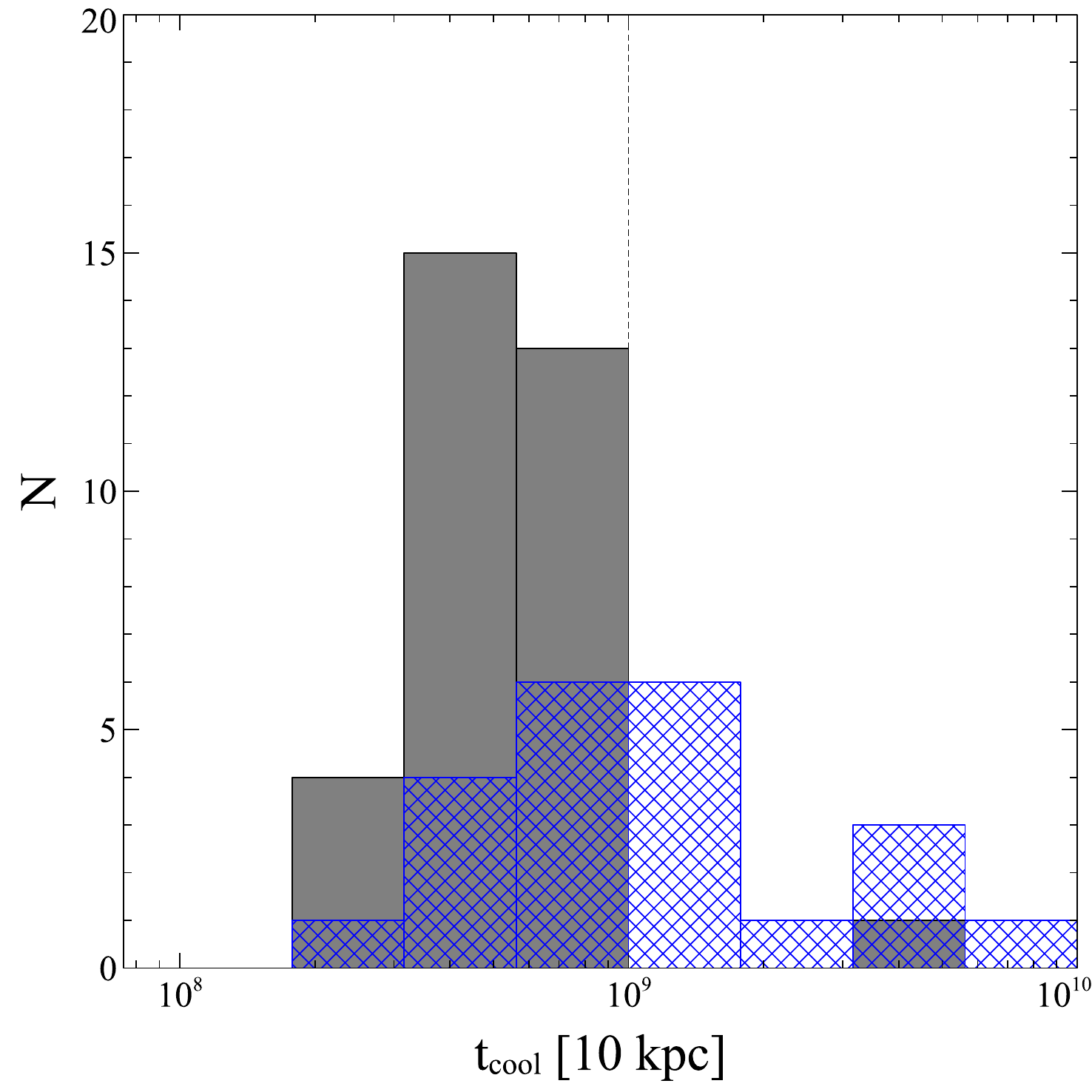}
\includegraphics[width=0.49\textwidth]{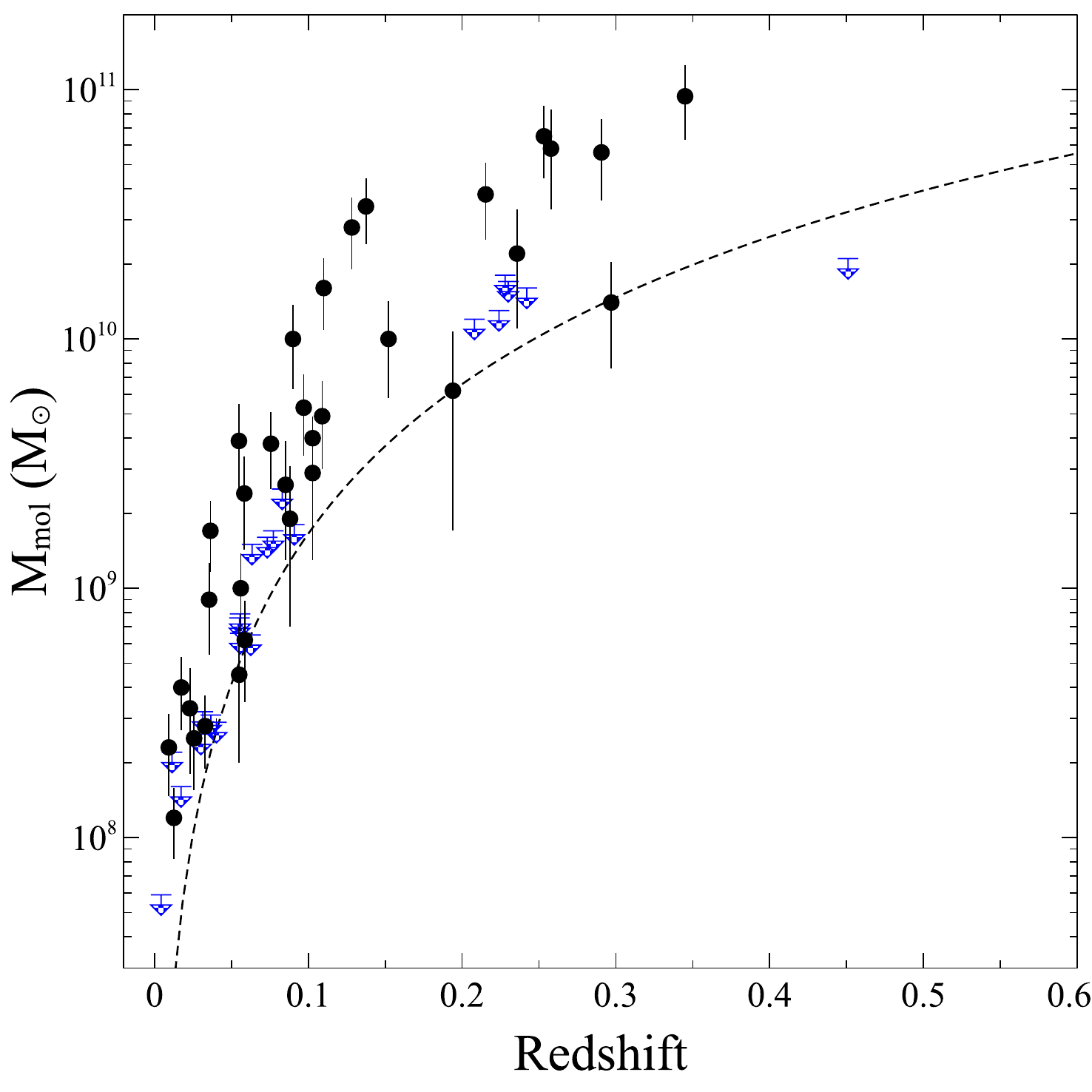}
\caption{\textit{Left:} Histogram of the cooling times at 10 kpc. Grey bars denote clusters detected in CO while blue bars denote non-detection. \textit{Right:} Molecular gas mass vs. redshift. The dashed curve represents the molecular mass limit that can be derived from CO(1-0) observations with the IRAM 30m telescope assuming a typical 300 km s$^{-1}$ linewidth and main beam temperature detection limit of 0.5 mK. The molecular gas mass for the two systems below this curve were derived from CO(3-2) observations. Black symbols denote systems observed with CO emission while blue symbols denote upper limits. \label{fig:cooling_time_threshold_exceptions}}
\end{figure*}

The histogram of central cooling times (see the left panel of Figure \ref{fig:cooling_time_threshold_exceptions}) shows two classes of outliers in the cooling time plot: (1) The system Abell 1060 with long central cooling time yet with detectable levels of molecular gas, and (2) eleven systems with short central cooling times but only upper limits to their molecular gas masses. The latter imply that a short cooling time does not guarantee the detection of molecular gas via CO emission. We consider these exceptions in turn.

Consistent with our findings, Abell 1060 was previously classified a ``weak`` cool core \citep[1 Gyr $< \rm t_{\rm cool} <$ 7.7 Gyr,][]{Mittal2009}. Accordingly, its molecular gas mass $1.2 \times 10^{8} \pm 0.4\ \text{M}_{\odot}$ is the lowest of the sample. As expected this value lies at the low end of the range of molecular gas typically observed in central galaxies \citep[$\sim$$10^{8-11}\ \text{M}_{\text{mol}}$,][]{Edge2001,Salome2003}, but well into the regime of molecular gas observed in normal elliptical galaxies \citep[$\sim$$10^{7-9}\ \text{M}_{\text{mol}}$][]{Young2011}.  Abell 1060 may have accumulated its molecular gas through a merger or perhaps through atmospheric cooling at an earlier time when its atmosphere was denser or AGN feedback was less effective.  The origin of molecular gas in elliptical galaxies is poorly understood. It may originate from both external (from another galaxy) and internal (stellar mass loss) processes \citep{Young2011}. 

\begin{figure*}[t!]
\centering
\includegraphics[width=0.33\textwidth]{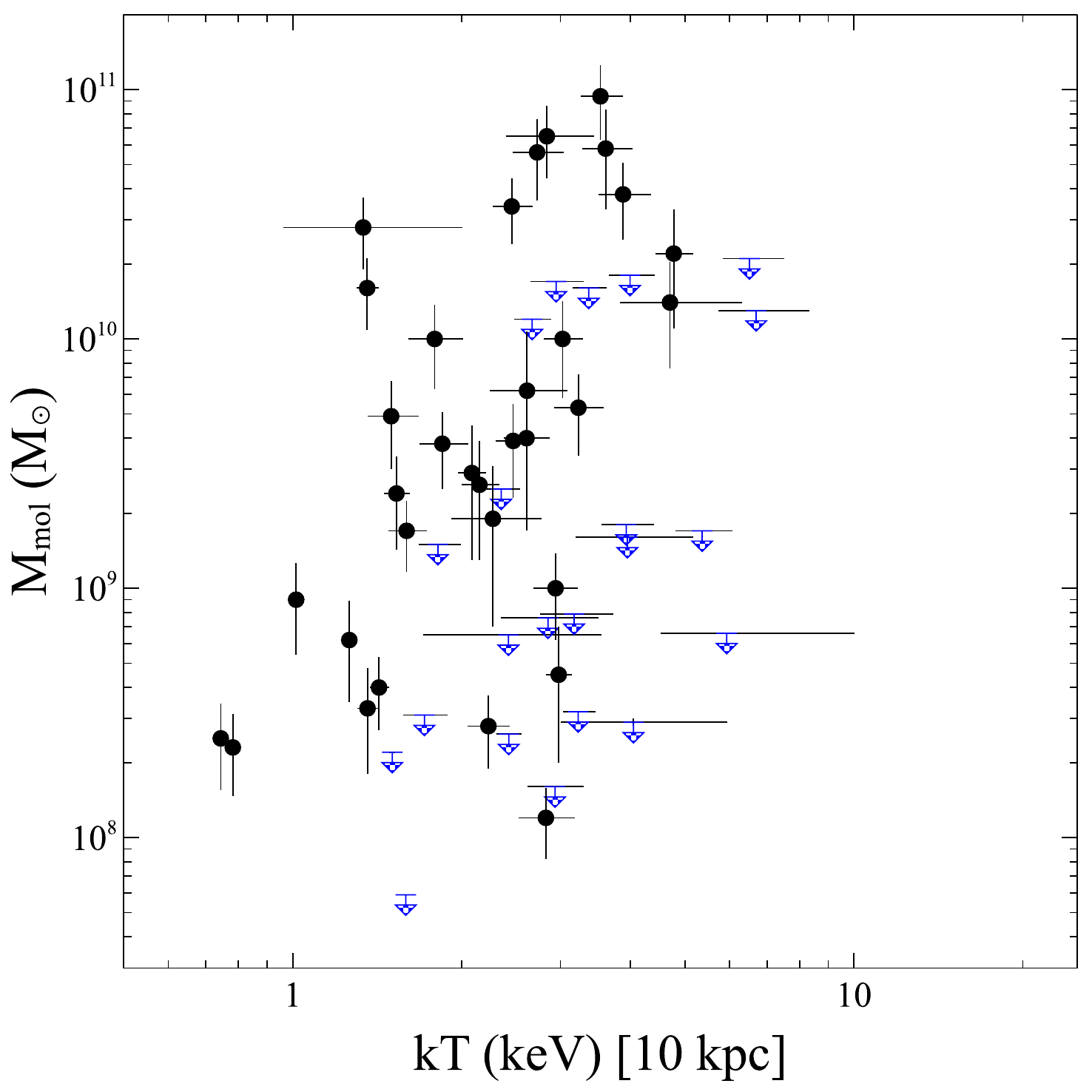}
\includegraphics[width=0.33\textwidth]{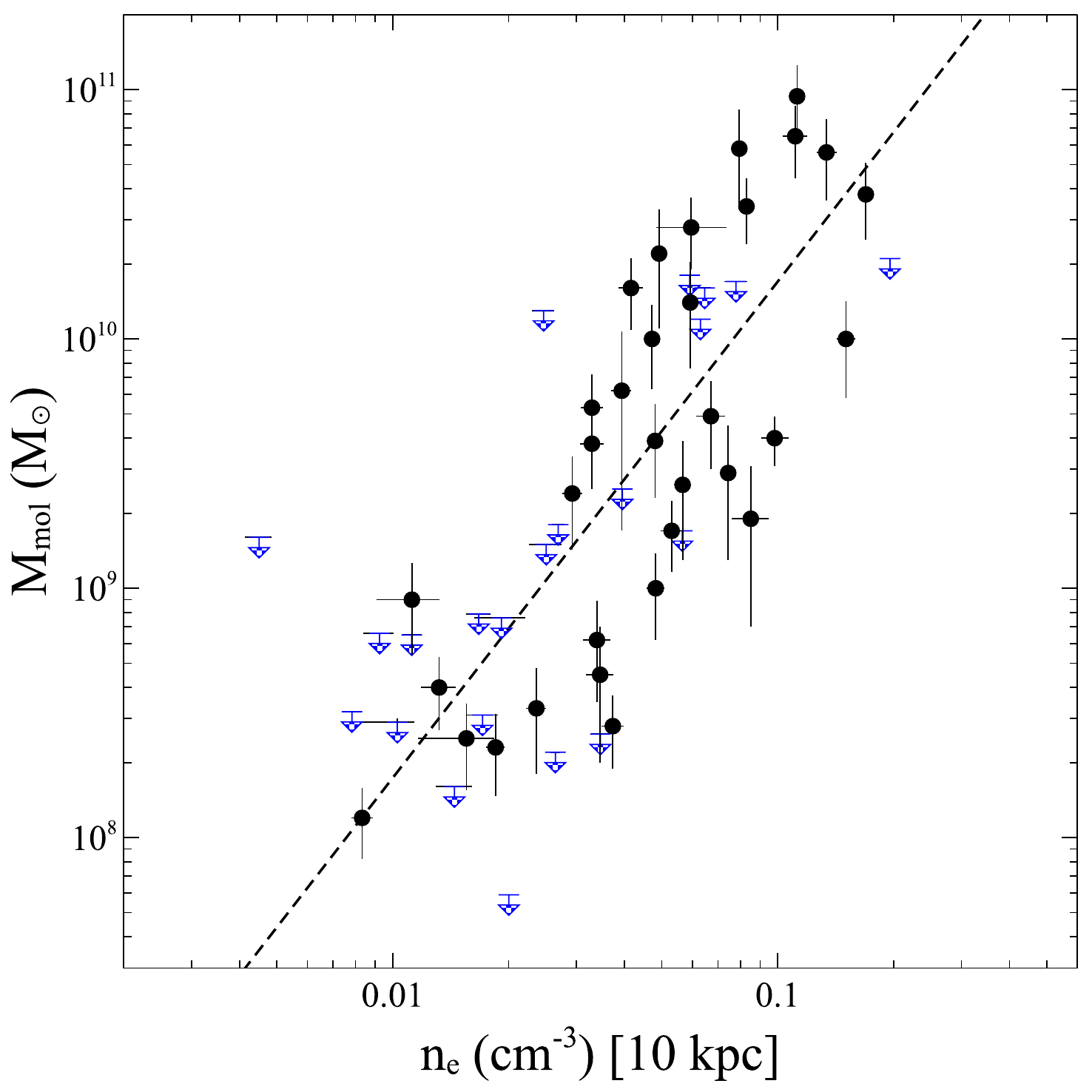}
\includegraphics[width=0.33\textwidth]{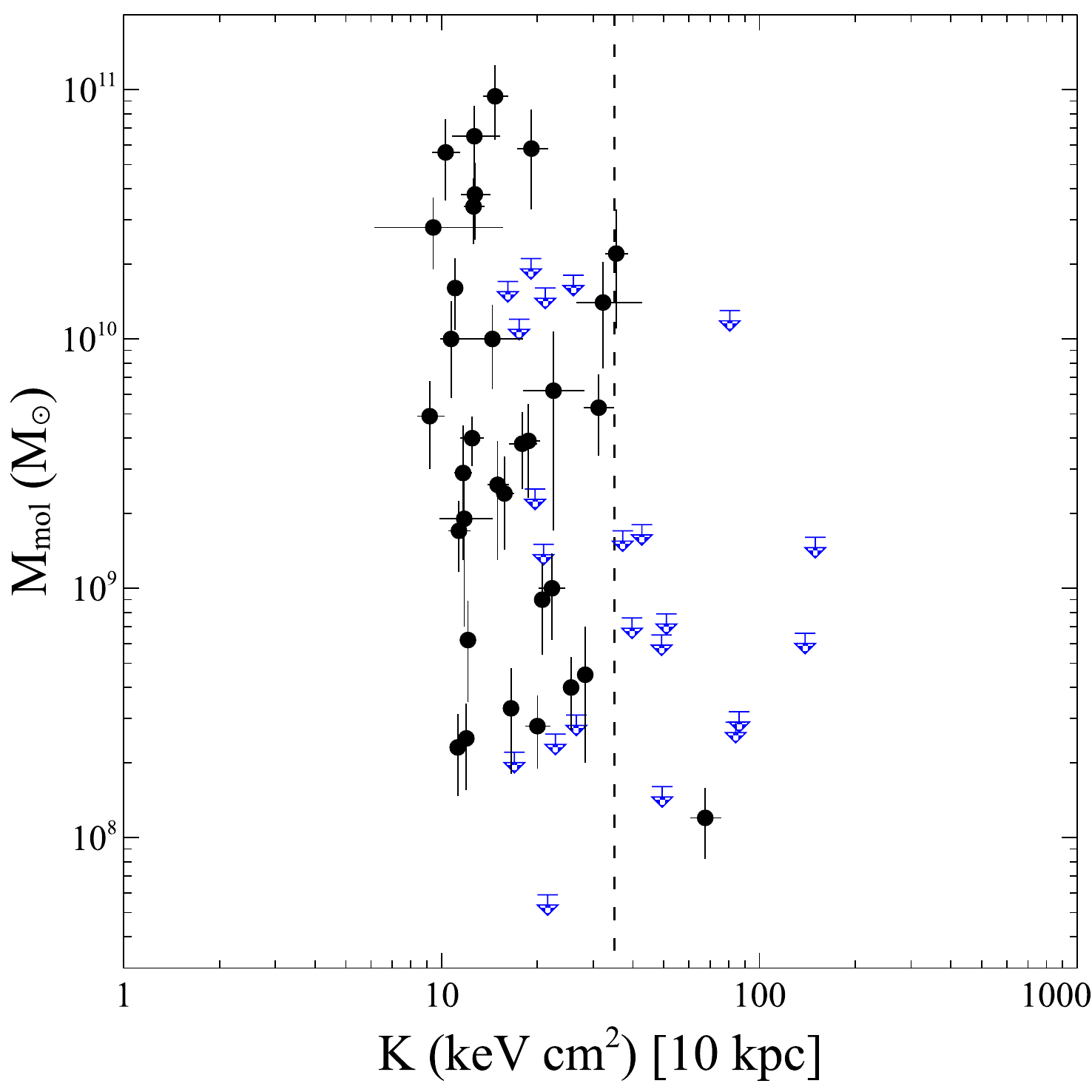}
\caption{Molecular gas mass vs. temperature, density, and entropy derived at 10 kpc. Black symbols denote systems observed with CO emission while blue symbols denote upper limits. In the middle panel, the dashed black line is a fit only to the systems with CO detection.}
\label{fig:xray_properties}
\end{figure*} 

Next, we address systems with short cooling times lacking CO detections. In the left panel of Figure \ref{fig:cooling_time_threshold_exceptions}, the detections (black) and non-detections (blue) occupy similar distributions in $\text{t}_{\text{cool}}$. A Kolmogorov-Smirnov (K-S) test gives a p-value of 0.22, which is too large to reject the null hypothesis that they were drawn from the same parent distribution. It is then possible and even likely that molecular gas is present but falls below the detection limit of the IRAM 30m observations. 

The right panel of Figure \ref{fig:cooling_time_threshold_exceptions} shows molecular gas mass plotted against redshift. The dashed line represents the molecular mass limit that can be achieved from our CO(1-0) observations with the IRAM telescope. Due to this detection limit, molecular clouds with CO(1-0) line emission below this curve will not be detectable. The distribution of upper limits is consistent with the detections, indicating that these objects likely contain substantial levels of molecular gas that lie below our detection limits. Alternatively, their molecular gas levels may be suppressed, perhaps due to AGN feedback. The connection between molecular gas and AGN feedback will be explored further in Section \ref{sub:molecular_gas_and_agn}. 


\subsection{Correlations Between Thermodynamic Parameters and Molecular Gas}\label{sub:x_ray_properties}

Understanding the relationship between the hot atmosphere and molecular gas is one of the goals of this paper. We have investigated the dependence of molecular gas mass with atmospheric temperature, density, and entropy in Figure \ref{fig:xray_properties}. The quantities are evaluated at 10 kpc to avoid resolution bias. While we find no obvious correlation between molecular gas mass and temperature, molecular gas mass and atmospheric density are strongly correlated. We find the best fit for the systems with CO detections only to be ($\text{R}^{2}$ = 0.60)
\begin{equation}
	\log{\text{M}_{\text{mol}}} = (1.99 \pm 0.51)\log{\text{n}_{\text{e}}} + (12.22 \pm 0.70)
\end{equation}
\noindent
The trend shows that molecular gas mass increases with atmospheric gas density, consistent with an atmospheric origin for the molecular gas. Discussed in more detail in Section \ref{sub:molecular_gas_and_agn}, higher ICM densities require greater heating to offset radiative cooling, hence a larger pool of molecular gas to feed the AGN. 


The trend between molecular gas mass and entropy ($\text{K} = \text{kTn}_{\text{e}}^{-2/3}$) shown in the third panel of Figure \ref{fig:xray_properties} shows that molecular gas is found in large quantities only in systems when the entropy falls below $\sim 35\ \text{keV}\ \text{cm}^{2}$.  The only outlier yet again is Abell 1060 as discussed earlier. Like cooling time, a low entropy atmosphere does not guarantee a detection of molecular gas.  Furthermore, a narrow spread is observed in the entropy distribution below $\sim$35 keV cm$^{2}$ with a mean of 17 keV cm$^{2}$ and standard deviation of 7 keV cm$^{2}$. These characteristics mirror those found for cooling time alone.

\subsection{Ratio of Cooling to Free-fall Time}\label{sub:cooling_freefall_ratio}

Atmospheric gas should become unstable to cooling when its ratio of the cooling to free-fall time ($\text{t}_{\text{cool}}/\text{t}_{\text{ff}}$) falls below $\sim$1 \citep{Nulsen1986,McCourt2012}. This criterion may rise above 10 in real atmospheres \citep{Sharma2012, Voit2015b, Gaspari2012, Gaspari2013}, which we investigate here.  Our analysis closely mirrors that of \citet{Hogan2017}, who adopted nebular emission rather than CO to trace cold gas.  Likewise our results and conclusions are similar. 

\begin{figure*}[b]
\centering
\includegraphics[width=0.49\textwidth]{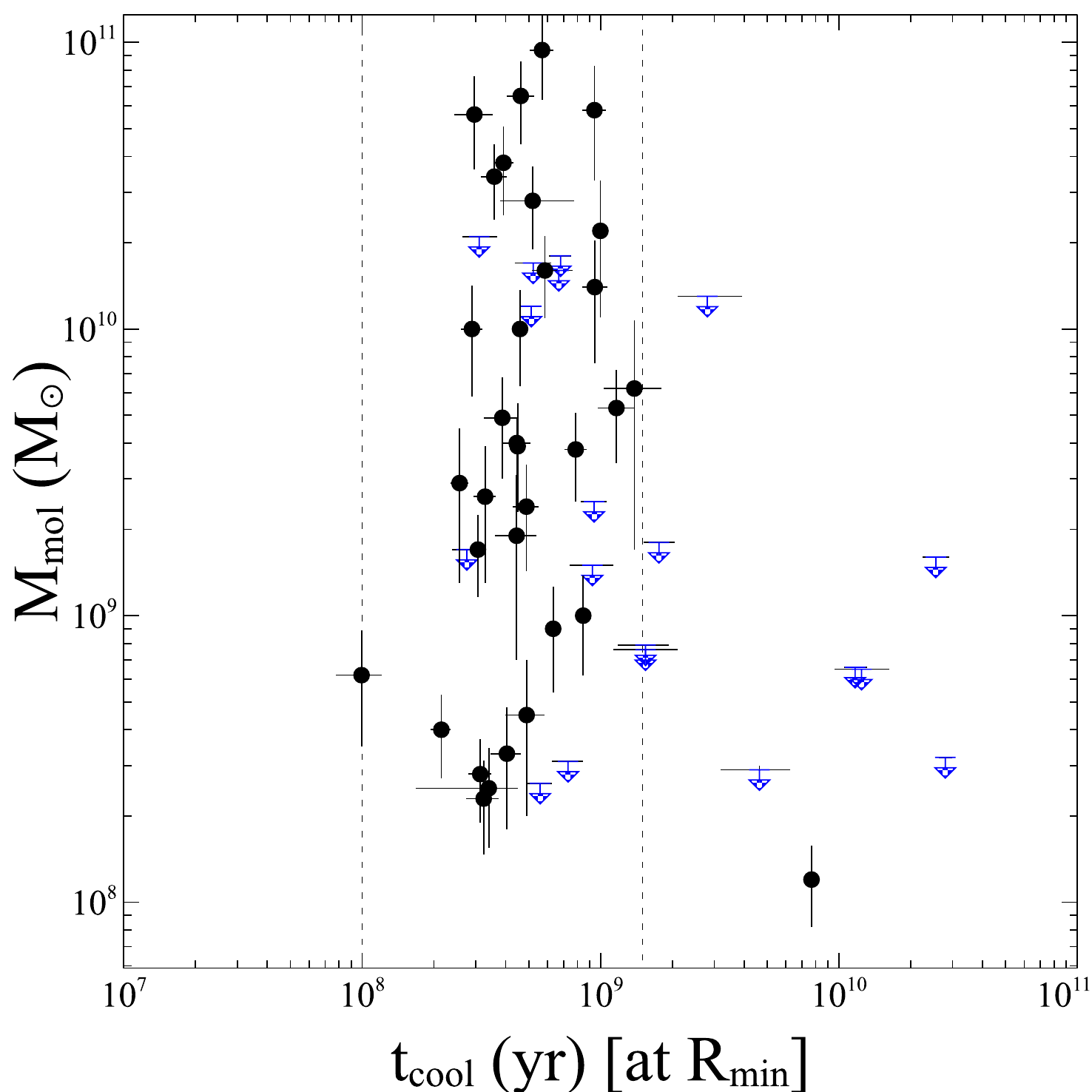}
\includegraphics[width=0.49\textwidth]{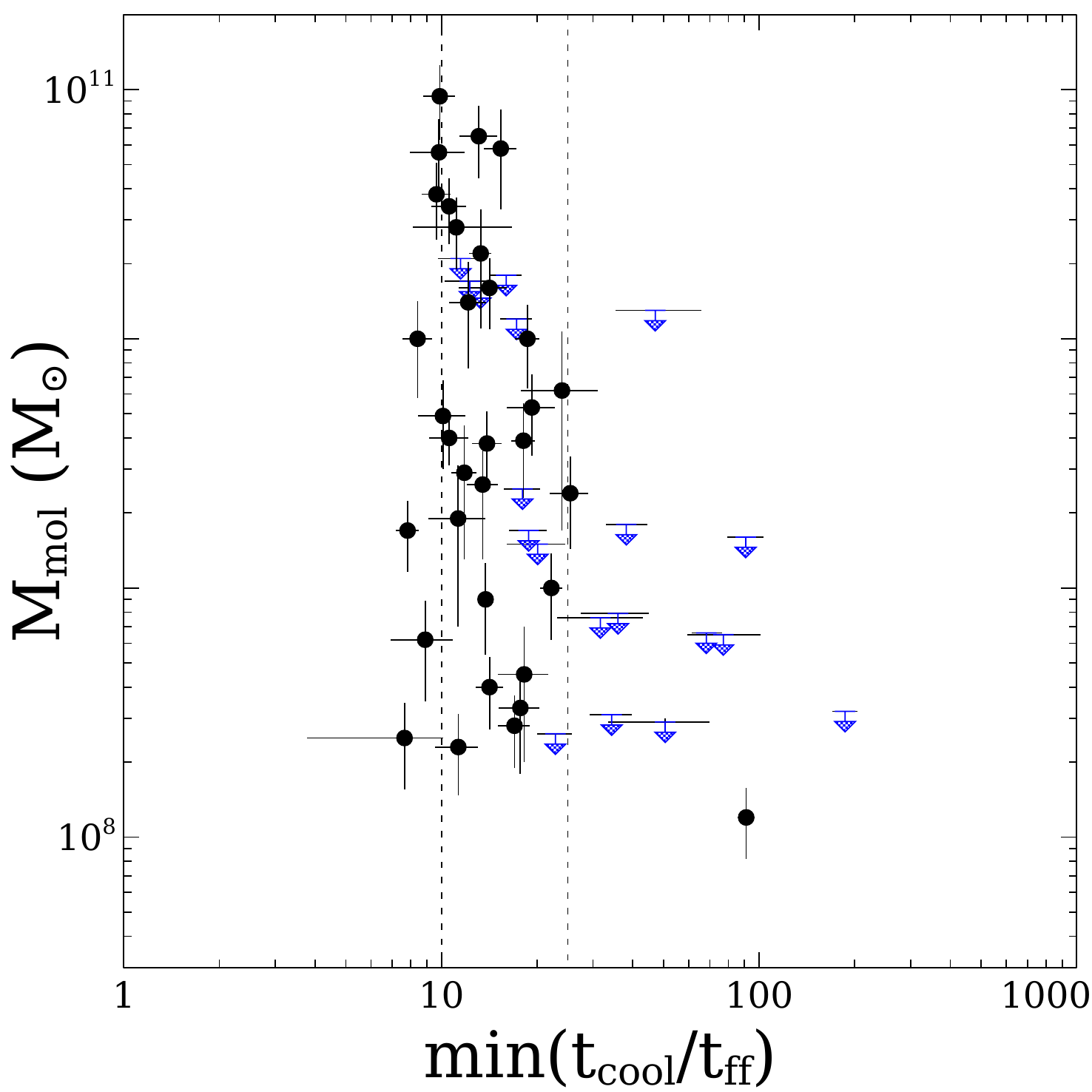}
\caption{\textit{Left panel}: Molecular gas mass vs. cooling time of the hot atmosphere at the radius which the minimum of cooling to free-fall time occurs. \textit{Right panel}:  Molecular gas mass vs. minimum cooling to free-fall time ratio.} 
\label{fig:co_molecular_gas_vs_ratio}
\end{figure*}

The left panel of Figure \ref{fig:co_molecular_gas_vs_ratio} shows molecular gas mass plotted against the cooling time of the atmosphere at the radius,  $\text{R}_{\text{min}}$, where the minimum ratio of $\text{t}_{\text{cool}}/\text{t}_{\text{ff}}$ is found.  The spread of cooling time for objects detected with molecular gas has a mean $\mu \simeq 0.46$ Gyr and standard deviation $\sigma \simeq 0.19$ Gyr. 

The right panel is similar but the cooling time is divided by the free-fall time at $\text{R}_{\text{min}}$.  Objects with detectable levels of molecular gas lie in the interval $10 \lesssim \text{min}(\text{t}_{\text{cool}}/\text{t}_{\text{ff}}) \lesssim 25$. The cutoffs on the high and low end of this range are abrupt.  The discontinuity at 25 corresponds roughly to the 1 Gyr threshold in cooling time found here and in \citet{Rafferty2006} and \citet{Cavagnolo2008}. The lower cutoff at 10 is similar to that found by \citet{Hogan2017} and \citet{Voit2015a}.  
Two to four points fall just below 10 in this figure but not significantly below when noise in the $\text{t}_{\text{cool}}/\text{t}_{\text{ff}}$ profile is taken into account (\citet{Hogan2017}, but see \citet{Prasad2015}.)

With a mean of 13.8 and a standard deviation of 4.6, the distribution of $\text{t}_\text{cool}/\text{t}_{\text{ff}}< 25$ appears narrower than the distribution of $\text{t}_{\text{cool}}$ alone. This narrowing suggests that that the free-fall time is contributing an observable and perhaps physical effect on the onset of thermally unstable cooling.  

\begin{figure*}[ht!]
\centering
\includegraphics[width=0.49\textwidth]{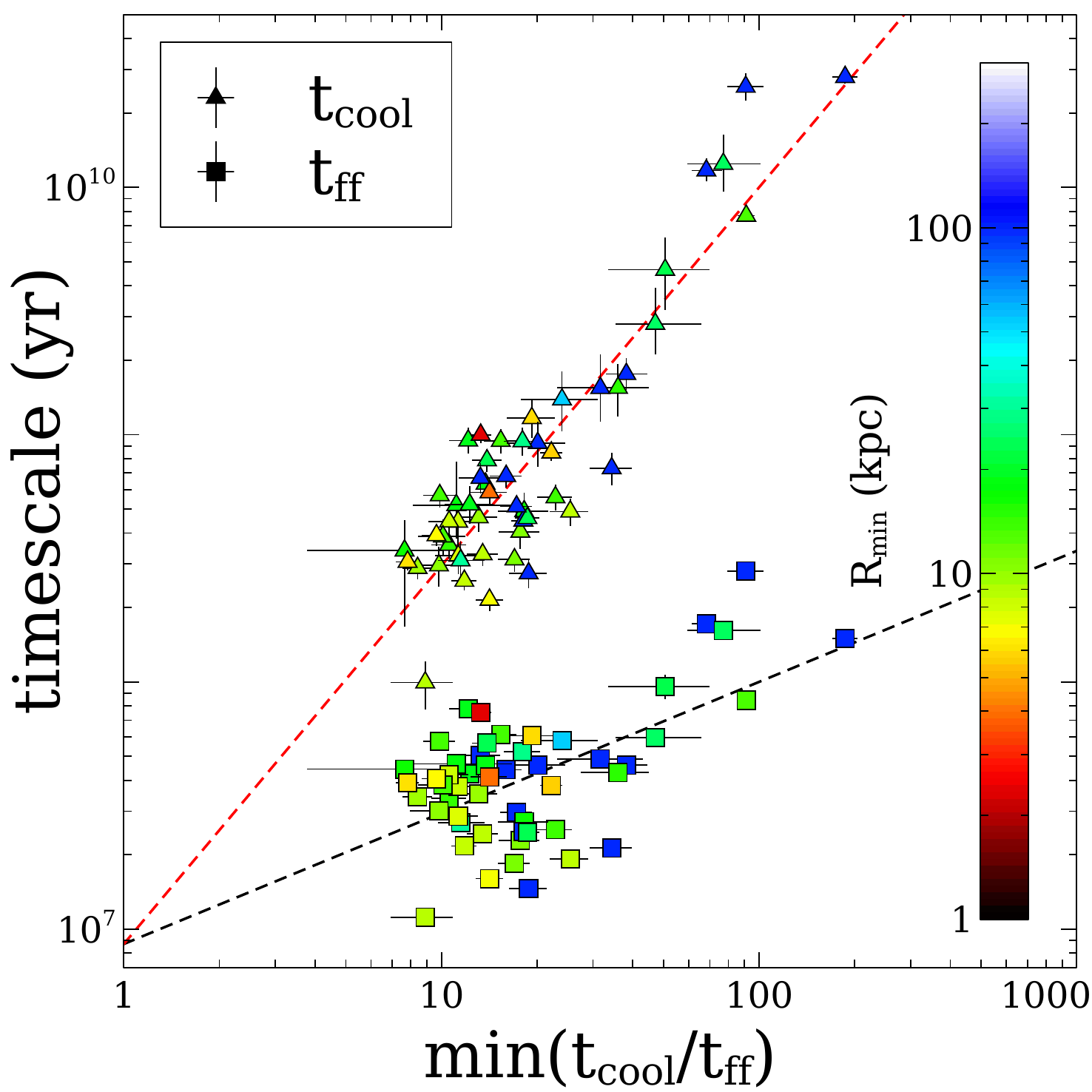}
\includegraphics[width=0.49\textwidth]{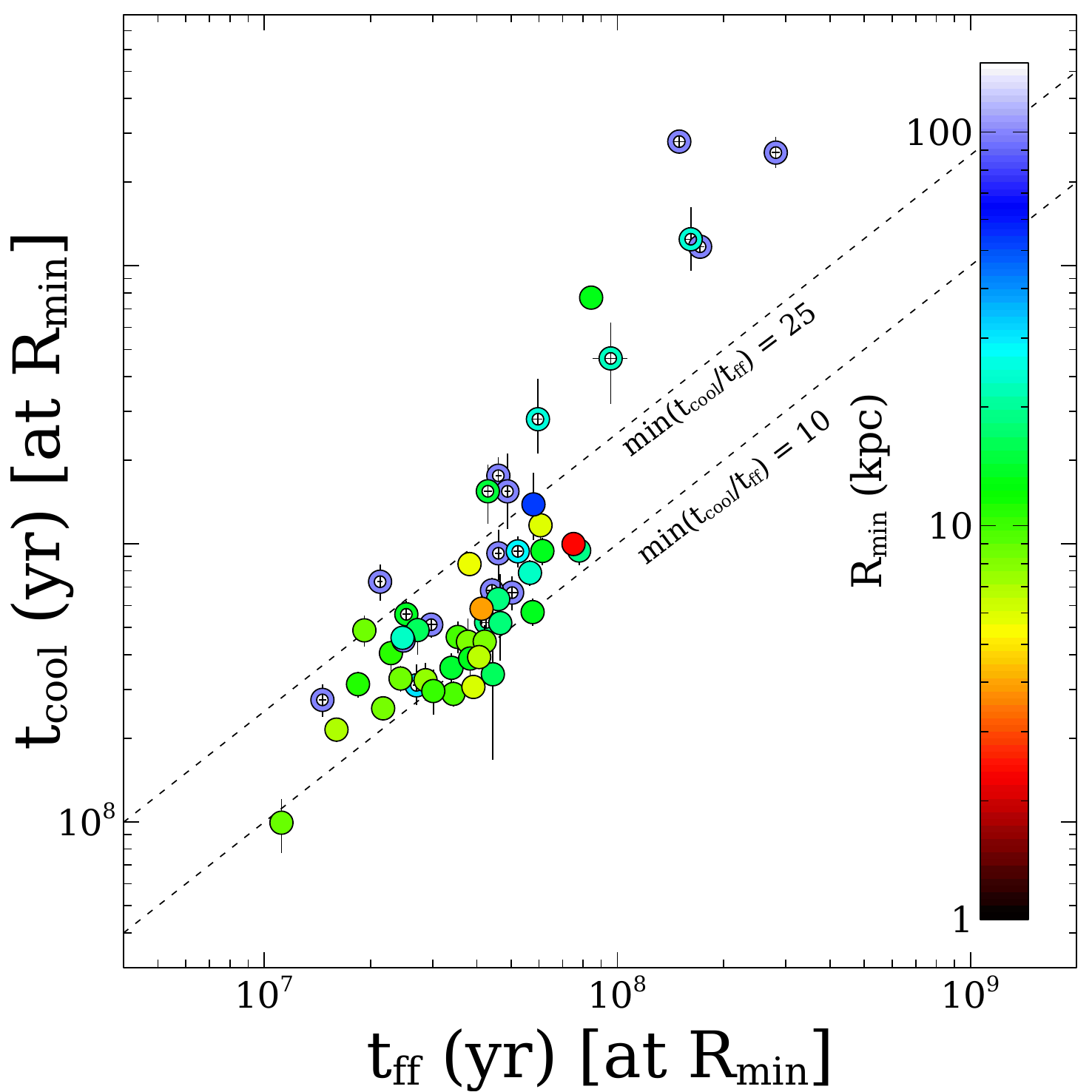}

\caption{\textit{Left panel}: Minimum cooling to free-fall time plotted against its numerator and denominator for our sample. The red and black dashed lines are linear fit in log-log space to the numerator and denominator against the ratio, respectively. \textit{Right panel}: The numerator plotted against denominator of the $\text{min}(\text{t}_{\text{cool}}/\text{t}_{\text{ff}})$ ratio. In both plots, the points are color-coded by the radius at which the minimum of the $\text{min}(\text{t}_{\text{cool}}/\text{t}_{\text{ff}})$ occurs.}
\label{fig:co_cooling_freefall_ratio}
\end{figure*}

However, the narrow spread of $\text{min}(\text{t}_{\text{cool}}/\text{t}_{\text{ff}})$, found also by \citet{Hogan2017}, can be attributed to a resolution bias. To demonstrate this bias, the minimum value of the ratio is plotted against its numerator and denominator in Figure \ref{fig:co_cooling_freefall_ratio}. The points are color coded by the value of $\text{R}_{\text{min}}$. 

This figure reveals two key points: First, $\text{min}(\text{t}_{\text{cool}}/\text{t}_{\text{ff}})$ is positively correlated with its numerator and denominator. The coefficients of determination, $\text{R}^2$, are 0.82 and 0.35, respectively. The stronger correlation with cooling time is evident over the entire range of $\text{min}(\text{t}_{\text{cool}}/\text{t}_{\text{ff}})$.  Conversely, the correlation with free-fall time vanishes for $\text{min}(\text{t}_{\text{cool}}/\text{t}_{\text{ff}}) < 35$, where cooling is strongest.  These trends show that $\text{t}_{\text{cool}}$ is primarily determining the $\text{min}(\text{t}_{\text{cool}}/\text{t}_{\text{ff}})$ ratio.  \citet{McNamara2016} and \citet{Hogan2017} reached the same conclusion.   

Secondly, the measured value of $\text{min}(\text{t}_{\text{cool}}/\text{t}_{\text{ff}})$ correlates with $\text{R}_{\text{min}}$. This trend is seen clearly in right panel of Figure \ref{fig:co_cooling_freefall_ratio} where we plot the numerator against the denominator of $\text{min}(\text{t}_{\text{cool}}/\text{t}_{\text{ff}})$ with points color-coded by $\text{R}_{\text{min}}$. For a given value of $\text{min}(\text{t}_{\text{cool}}/\text{t}_{\text{ff}})$, a large numerator is offset by a large denominator and conversely so.  In other words, the narrow distribution of $\text{min}(\text{t}_{\text{cool}}/\text{t}_{\text{ff}})$ is plausibly explained by a resolution bias.  Shorter cooling times are alway measured closer to the nucleus where the free-fall time is likewise shorter.  The converse is also true.  The upshot is that $\text{min}(\text{t}_{\text{cool}}/\text{t}_{\text{ff}})$ is condemned to lie in a narrow range because of how it is defined.  A similar conclusion was reached by \citet{Hogan2017}. 

\cite{Hogan2017} also considered the possibility that the minimum in the $\text{t}_{\text{cool}}/\text{t}_{\text{ff}}$ profiles may actually be a floor, rather than a clear minimum with an upturn at smaller radii. When the mass profile is approximately isothermal and the entropy profile follows a power-law slope of $\text{K} \propto \text{r}^{2/3}$ in the inner region,  $\text{t}_{\text{cool}}/\text{t}_{\text{ff}} \propto 1/[\Lambda (\text{kT})^{1/2}]$  is found \citep{Hogan2017}. This expression is independent of radius suggesting that the upturns observed in $\text{t}_{\text{cool}}/\text{t}_{\text{ff}}$ profiles are produced by density inhomogeneities along the line of sight \citep{Hogan2017}. 

In summary, we preferentially observe objects with molecular gas when the cooling to free-fall time ratio lies in the narrow range $10 \lesssim \text{min}(\text{t}_{\text{cool}}/\text{t}_{\text{ff}}) \lesssim 25$. The upper bound corresponds to the $\sim1$ Gyr cooling time threshold; no object falls significantly below 10.  The narrow spread can be attributed to resolution bias, although a physical origin cannot be ruled out.

\begin{figure*}[ht!]
\centering
\includegraphics[width=0.49\textwidth]{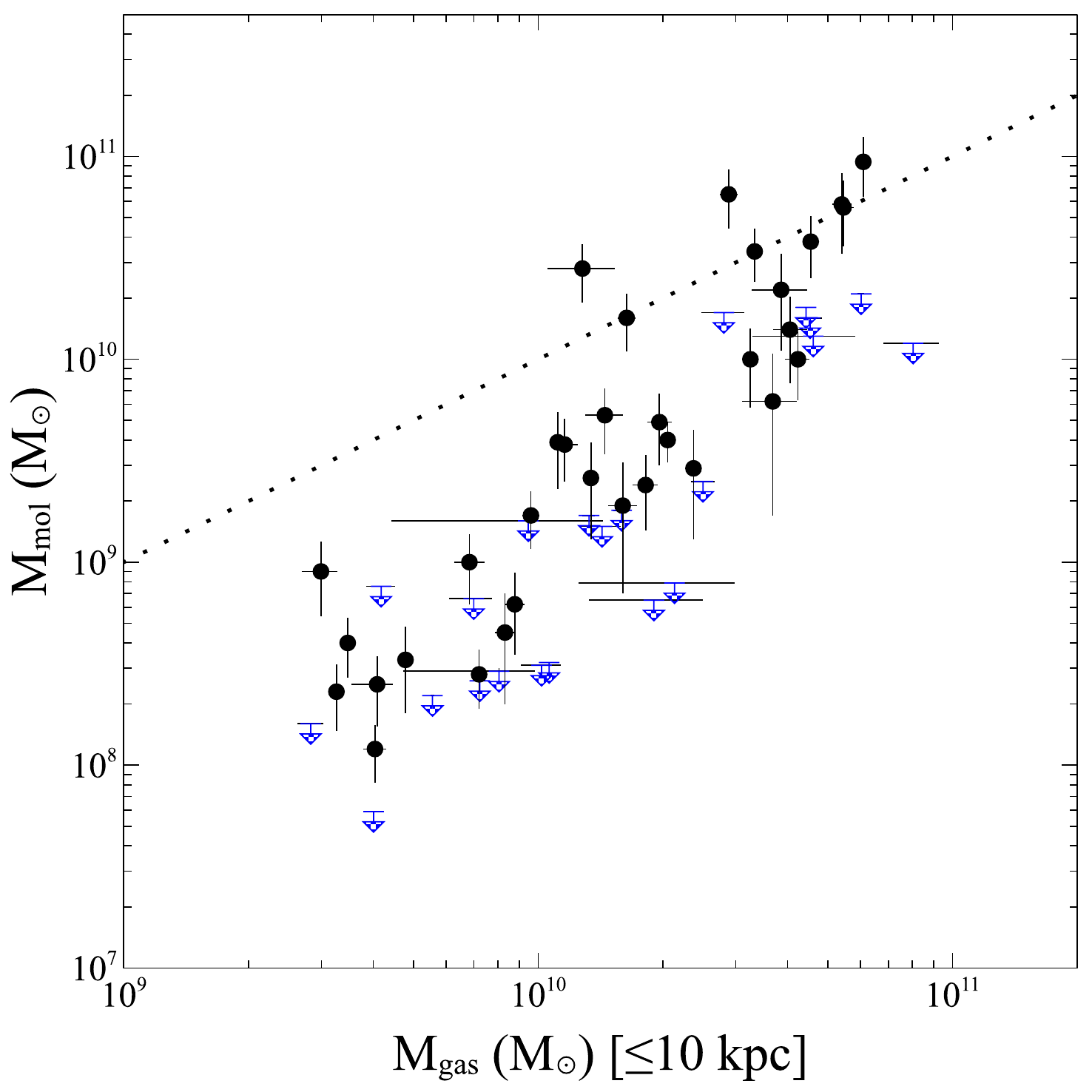}
\includegraphics[width=0.49\textwidth]{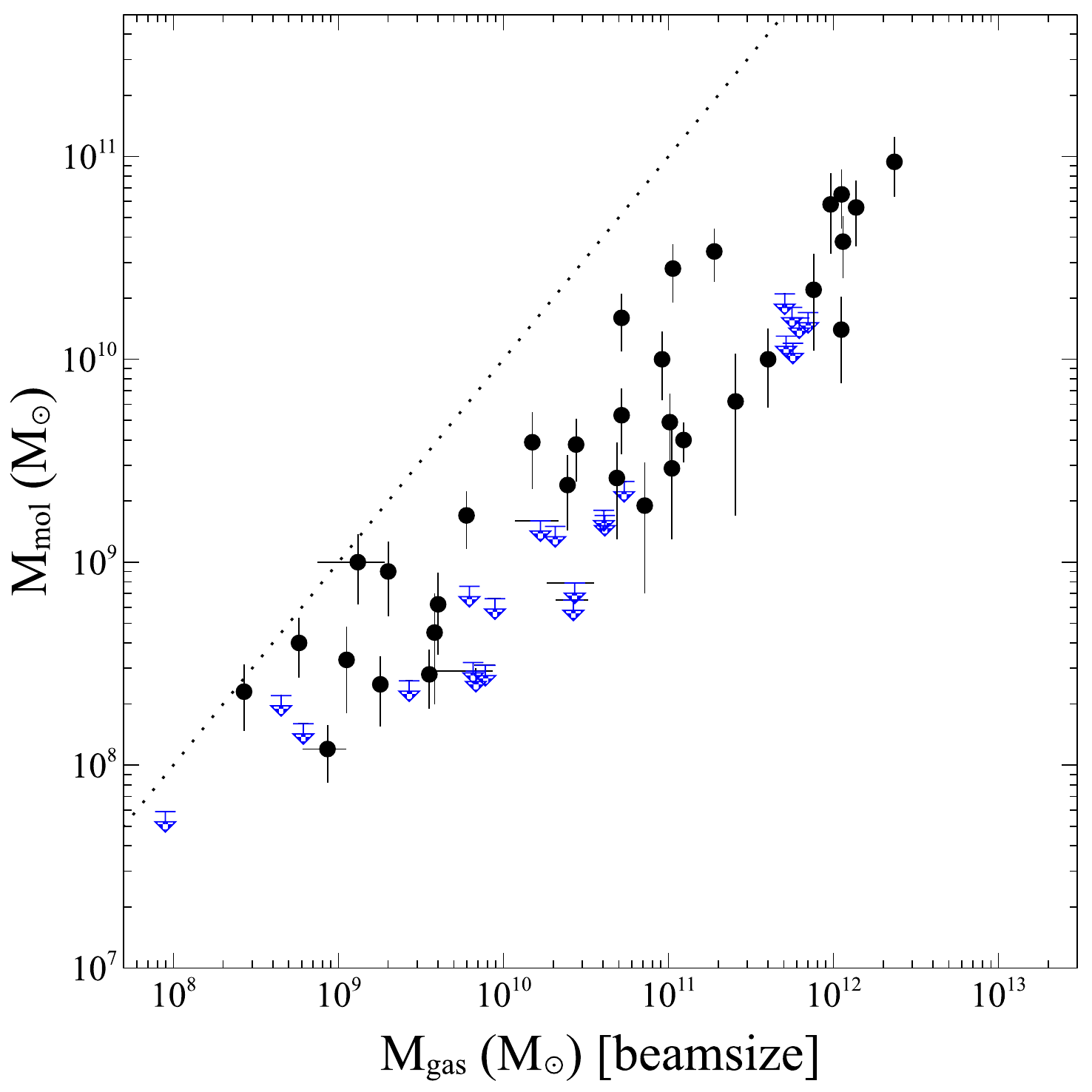}
\caption{{\it Left panel:} Molecular gas mass enclosed within the IRAM 30m beam vs. Hot gas mass enclosed within 10 kpc in radius from the cluster core. {\it Right panel:} Molecular gas mass vs. Hot gas mass both enclosed within the IRAM 30m beam. The dotted line shows equality of the quantities. Black symbols denote systems observed with CO emission while blue symbols denote upper limits.}
\label{fig:molecular_and_hot_gas_mass}
\end{figure*}

\subsection{Does Thermally Unstable Cooling Ensue When $\text{t}_{\text{cool}}/\text{t}_{\text{ff}}\lesssim 10$?}\label{sub:precipitation_models}

Thermally unstable cooling in a stratified, plane-parallel atmosphere is thought to occur when the ratio of the local cooling time to free-fall time, $\text{t}_{\text{cool}}/\text{t}_{\text{ff}}$, falls below unity  \citep{McCourt2012}.  Recent studies have suggested that this ratio rises above unity in realistic, three dimensional atmospheres \citep{McCourt2012,Singh2015,Gaspari2012,Li2015,Voit2015a,Voit2015b,Prasad2015,Lakhchaura2016}.  For example, \citet{Voit2015b} found that nebular emission became prevalent in systems where $4 \lesssim \text{min}(\text{t}_{\text{cool}}/\text{t}_{\text{ff}}) \lesssim 20$.  Numerical feedback simulations by \citet{Li2015} found that atmospheres become thermally unstable when $1 \lesssim \text{min}(\text{t}_{\text{cool}}/\text{t}_{\text{ff}}) \lesssim 25$.  These studies suggest that those systems with $\text{t}_{\text{cool}}/\text{t}_{\text{ff}}$ lying below 10 are rapidly cooling into molecular clouds, while the atmospheres of those lying above 10 are stabilized by AGN feedback \citep{Voit2016, Gaspari2012}.   


The measurements shown in Figure \ref{fig:co_molecular_gas_vs_ratio} are broadly consistent with this statement.  There is little doubt that the molecular clouds formed from thermally unstable cooling.  But significant and consequential issues remain:
\begin{itemize}

\item First, the cycling of heating and cooling that may lead to large fluctuations in central atmospheric gas density and cooling time is not observed here or by \cite{Hogan2017}.  Because the gravitational potentials are fixed, only the cooling time may be varying.  Our data show the cooling time varies much less than contemplated by \citet{Li2015} and \citet{Voit2016}.  For example, over the course of their 6 Gyr simulation, \cite{Li2015} found that this ratio falls below 10 about one-fifth of the time.   We would then expect that of the 55 clusters observed here, roughly 10 should lie below $\text{min}(\text{t}_{\text{cool}}/\text{t}_{\text{ff}}) < 10$.  This number would increase to 20 were we to include the \citet{Hogan2017} sample.  Only two to four objects fall just below 10, and no object falls significantly below.  

This discrepancy is unlikely due to sampling bias. We have explored a wide range of AGN power ($10^{42-46}\ \text{erg}\ \text{s}^{-1}$), halo mass ($\text{M}_{2500}\sim 10^{13-14}\ \text{M}_{\odot}$), molecular gas mass ($10^{8-11}\ \text{M}_{\odot}$), and redshift ($z\sim 0-0.4$). We should be sampling the entire feedback cycle of heating and cooling contemplated by \citet{Li2015} and others. 

The upshot is that our measurements are in tension with models that assume molecular gas forms from local thermal instability growing from linear density perturbations \citep{McCourt2012, Sharma2012}.   Figure \ref{fig:co_molecular_gas_vs_ratio} shows that systems rich in molecular gas lie in the range $10<\text{t}_{\text{cool}}/\text{t}_{\text{ff}}<25$, the regime where atmospheres are expected to be thermally stable \citep{Gaspari2012}.  Nevertheless,  ALMA observations of many systems indicate that molecular gas is condensing currently on a timescale shorter than the free-fall timescale.  

Many cluster simulations show large temperature, density, and entropy fluctuations in response to AGN activity that damp away over several tens to hundreds
of Myr.  However, in real clusters, feedback is unrelenting.  Cluster atmospheres would not settle before the next significant AGN outburst.   Feedback is a gentle process. Host atmospheres remain remarkably stable throughout the feedback cycle and over large variations in jet power \citep{McNamara2016, Hogan2017}.  Simulations have begun to incorporate this into feedback models \citep{Sijacki2007, Gaspari2017}.  Solving this problem will surely lead to new insights into the subtlety of self regulation.

\begin{figure*}[ht!]
\centering
\includegraphics[width=0.49\textwidth]{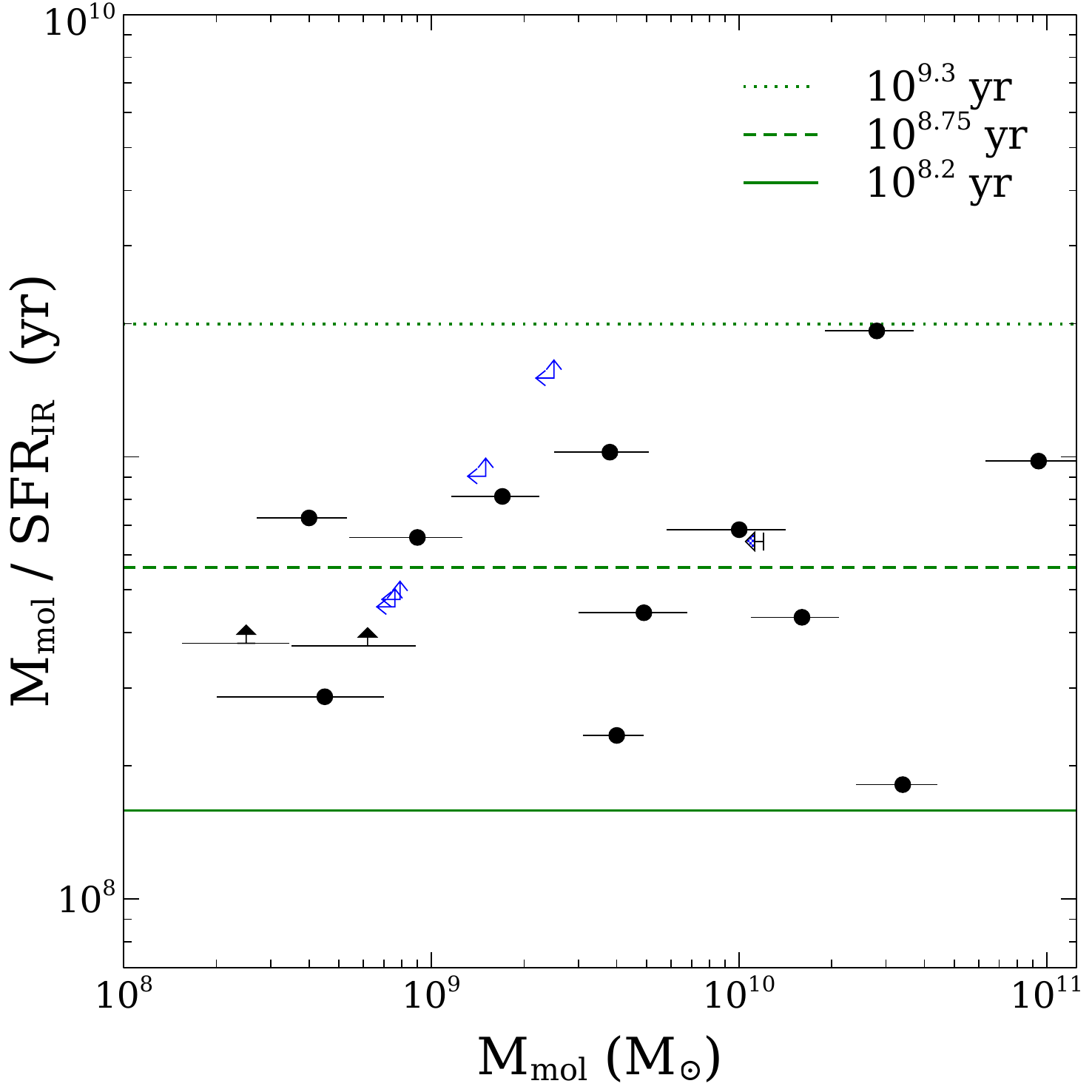}
\includegraphics[width=0.49\textwidth]{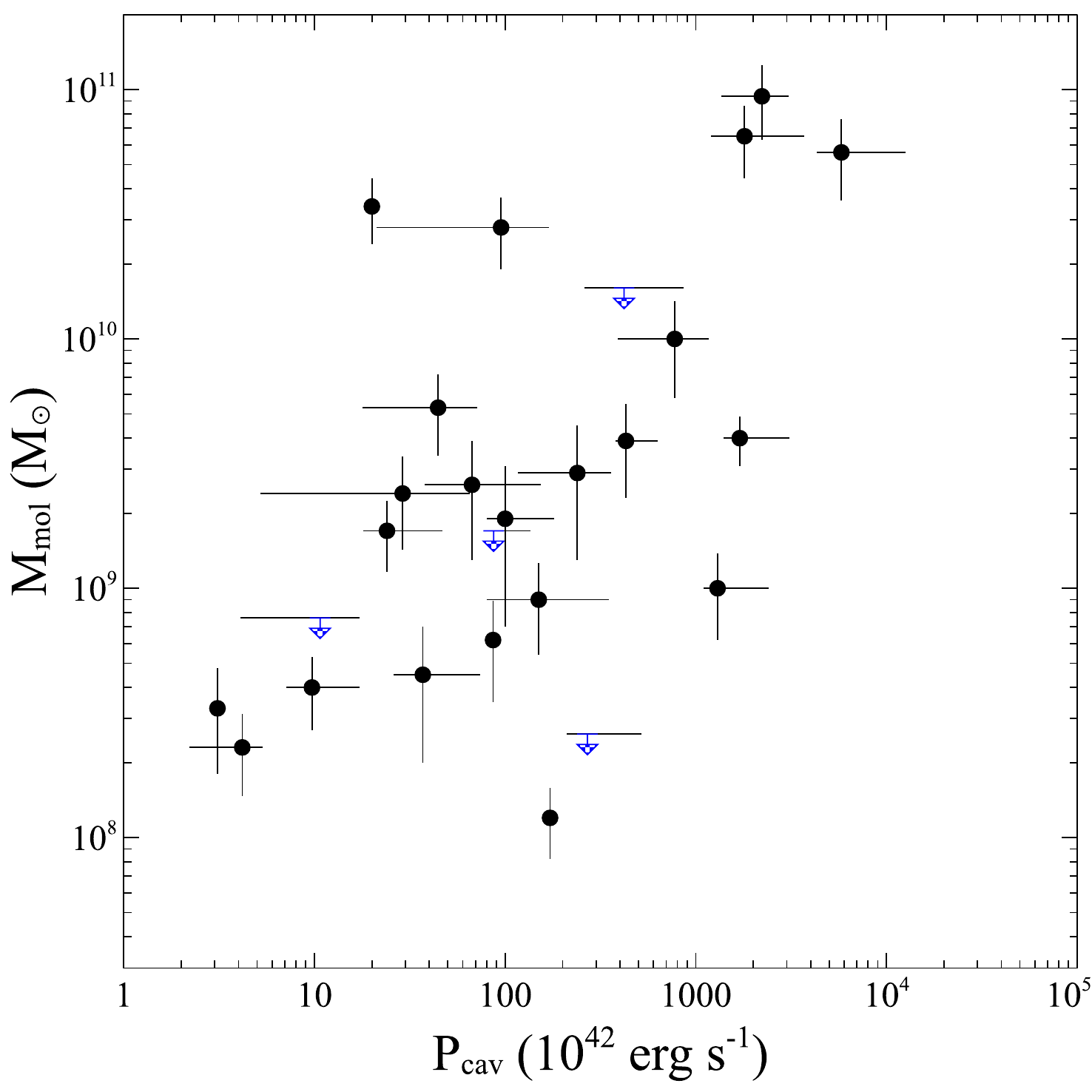}
\caption{{\it Left panel:} Depletion timescale of molecular gas due to star formation ($\text{M}_{\text{mol}}/\text{SFR}_{\text{IR}}$) vs. molecular gas mass. Star formation rates were taken from \cite{Odea2008}. {\it Right panel:} Molecular gas mass vs. Cavity power. Black symbols denote systems observed with CO emission while blue symbols denote upper limits.}
\label{fig:cavity_power_and_molecular_gas}
\end{figure*}





\item 
All systems with molecular gas detections lie in the narrow range $10 \lesssim \text{min}(\text{t}_{\text{cool}}/\text{t}_{\text{ff}}) \lesssim 25$.  We find no indication that atmospheres hosting systems with the largest molecular gas reservoirs lie preferentially at the low end, and presumably more thermally unstable end, of $\text{min}(\text{t}_{\text{cool}}/\text{t}_{\text{ff}})$.  The most powerful jets are not associated with larger molecular gas reservoirs (Fig. 9), nor does jet power correlate with higher values of $\text{min}(\text{t}_{\text{cool}}/\text{t}_{\text{ff}})$ (Figure \ref{fig:ratio_vs_pcav}). 

The most general observation that can be made is that thermally unstable cooling ensues when the cooling time of the atmosphere drops below $\sim 1$ Gyr, or $\text{min}(\text{t}_{\text{cool}}/\text{t}_{\text{ff}})\lesssim 25$, or $K \lesssim 35~\rm keV~cm^2$.  Measurements provide no further indication of the role the free-fall time plays, as the value of the ratio is driven almost entirely by the cooling time \citep{McNamara2016, Hogan2017}.

\item Apart from one object, Hydra A \citep{Hamer2014}, ALMA observations have found little evidence for large-scale molecular gas disks (\citet{Russell2017} and references cited there) envisioned by the cold chaotic accretion model \citep{Gaspari2012} or the precipitation model \citep{Li2015}.    

\item Finally,  the \citet{Voit2016} precipitation model assumes cluster entropy profiles become isentropic (constant entropy) in their centers,
where  $\text{t}_{\text{cool}}/\text{t}_{\text{ff}} \lesssim 10$.   In this model, the absence of an entropy gradient in the isentropic core promotes thermally
unstable cooling without the aid of uplift from the radio AGN.  Isentropic cores are rare.  \citet{Panagoulia2014b} and \citet{Hogan2017} found no evidence for isentropic cores in
clusters.  Instead, the entropy profiles scale approximately as $K\propto r^{2/3}$ in the inner few tens of kpc.  \citet{Hogan2017} pointed out that this scaling naturally leads to
a  $\text{t}_{\text{cool}}/\text{t}_{\text{ff}}$ profile that reaches a constant in the inner few tens of kpc. This is certainly the region most prone to thermally
unstable cooling. 
Precipitation in non-isentropic cores may be stimulated by uplift  \citep{Voit2016}, which is consistent with ALMA observations.

\end{itemize}

\section{Discussion} \label{sec:discussion}

\subsection{Cooling Out of the Hot Atmosphere as a Plausible Origin of Cold Gas}\label{sub:thermal_instability}

Molecular gas in early-type galaxies rarely rises above $\sim 10^8 ~\rm M_\odot$ \citep{Young2011}.  However, many central cluster galaxies are exceptionally rich in molecular gas, with masses approaching $\sim 10^{11} ~\rm M_\odot$ in some instances. Studies have shown that nebular emission and star formation in central galaxies are correlated with short central cooling times of the hot intracluster medium. For example, systems with H$\alpha$ emission and blue continuum emission have been preferentially detected with central cooling times $\lesssim$ 1 Gyr and entropy index $\lesssim$ 30 keV cm$^{2}$ \citep{Cavagnolo2008,Rafferty2008,Voit2015b}. While H$\alpha$ and blue light trace the cold $10^{4}$ K ionized gas and recent star formation, respectively, CO emission probes the gas directly fueling star formation and AGN.  Molecular gas drives galaxy and black hole co-evolution. 
If the molecular clouds have indeed cooled from hot atmospheres, mass continuity dictates that their hot reservoirs must be significanly more massive than the molecular reservoirs in central galaxies. 

On the left panel in Figure \ref{fig:molecular_and_hot_gas_mass}, we plot molecular gas mass against the surrounding atmospheric mass observed in the inner 10 kpc, where most of the molecular gas is seen by ALMA.  Two features are seen in this plot.  First, the atmospheric mass is positively correlated with the molecular gas mass.  Second, in most instances the atmospheric mass exceeds the molecular gas mass. Taking the atmospheric gas mass within the beam (right panel of Figure \ref{fig:molecular_and_hot_gas_mass}), the average fraction of cold to hot gas within the beam is 0.18 on average. This trend is consistent with the density trend shown in Figure \ref{fig:xray_properties}. Both are plausible with the molecular gas having cooled from the hot atmosphere. In some cases, such as Abell 1835, the molecular gas mass exceeds the atmospheric mass in its vicinity. The single dish beam of IRAM for A1835 is roughly 104 kpc and measures a molecular to atmospheric gas ratio of $6.5\times10^{10}\ \text{M}_{\odot}\ /\ 1.1\times10^{12}\ \text{M}_{\odot} \approx 0.06$. However, within 10 kpc of its BCG, ALMA observations measures this ratio to be $5\times10^{10}\ \text{M}_{\odot}\ /\ 1.12\times10^{10}\ \text{M}_{\odot} \approx 4.46$. In the few cases where this is true, the gas likely cooled from larger radii or has lingered over time. On the whole, these trends would be difficult to explain by an external origin, such as a merger or stripping from a plunging galaxy.

The left panel of Figure \ref{fig:cavity_power_and_molecular_gas} plots the depletion timescale for star formation to consume the molecular gas. The median depletion timescale is $\sim$1 Gyr (see also \cite{Odea2008}). Taken at face value, the depletion timescales indicated long-lived star formation.  Long-lived star formation is in tension with the molecular gas kinematics, discussed below.

.
\subsection{A Relationship Between Molecular Gas and AGN Feedback?}\label{sub:molecular_gas_and_agn}

AGN power roughly balances energetically the radiative losses from the atmospheres (e.g. \cite{Birzan2004}). This balance can be maintained only if accretion onto the black hole responds promptly to a change of atmospheric state. Accretion from the hot atmosphere surrounding the nuclear black hole should occur at some level \citep{Allen2006}. Bondi accretion alone would be insufficient to fuel the most powerful AGN if their nuclear black hole masses follow the $M-\sigma$ relation \citep{Rafferty2006, McNamara2011, Hardcastle2007, Narayan2011}. On the other hand, molecular gas is abundant and could easily supply the fuel, found here and elsewhere \citep{Pizzolato2005, Gaspari2012, Li2014a, Prasad2015}. 

Assuming hot atmospheres are stabilized by AGN feedback, higher atmospheric gas densities would require higher heating levels.  Furthermore, systems with higher atmospheric gas densities would likewise require larger molecular gas reservoirs to fuel AGN (middle panel of Figure \ref{fig:xray_properties}). 
We would then expect molecular gas mass to correlate with AGN power. 

The right panel of Figure \ref{fig:cavity_power_and_molecular_gas} reveals no strong correlation between these quantities. We instead observe a three decade scatter in AGN power for a given molecular gas mass. Taken at face value, this figure is inconsistent with molecular gas fueling of AGN feedback.  However, in a similar analysis, \citet{McNamara2011} pointed out that only a small fraction of the molecular gas near the nucleus would be required to fuel the AGN at the observed power levels. The single dish masses given here are sensitive to molecular gas spread over tens of kiloparsecs.  Therefore these measurements do not have the spatial sensitivity to reveal such a correlation should it exist. Nevertheless, such a correlation should exist and should be pursued with high resolution ALMA measurements. 



\subsection{Molecular Gas Stimulated by Uplift?}\label{sub:alma_observation}

Molecular gas maps of more than a half dozen systems show a surprising level of complexity \citep{Salome2003, David2014, McNamara2014, Russell2014, Russell2016b, Russell2016a, Vantyghem2016, Tremblay2016, Russell2017}. Apart from Hydra A \citep{Hamer2014}, ordered motion in disks or rings is largely absent. Instead the molecular gas lies in filaments and lumps that are often displaced from the center of the host galaxy. Their velocities and velocity dispersions are surprisingly low.  Molecular clouds in central galaxies are often moving well below both the free fall speed and the velocity dispersions of the stars.  Their low velocities and disordered motions suggest the clouds are young and have only recently cooled from the surrounding atmosphere. These properties are difficult to square with the long, $\sim 1$ Gyr depletion timescale to star formation seen in Figure \ref{fig:cavity_power_and_molecular_gas}. 

In many systems the molecular gas is projected behind buoyantly-rising X-ray bubbles. Examples include Perseus \citep{Salome2005, Lim2008}, Phoenix \citep{Russell2017}, Abell 1835 \citep{McNamara2014}, PKS 0745-091 \citep{Russell2016a}, and Abell 1795 \citep{Russell2017}. This phenomenology indicates that the molecular clouds are being tossed around by rising X-ray bubbles or molecular cloud condensation is initiated by cool atmospheric gas lifted to higher altitudes in the wakes of buoyantly rising X-ray bubbles \citep{Revaz2008, Li2014a, McNamara2014, Brighenti2015, McNamara2016, Voit2016}. Both mechanisms are occurring at some level. 


ALMA observations are providing clues to why hot atmospheres become thermally unstable when  $\text{min}(\text{t}_{\text{cool}}/\text{t}_{\text{ff}})$ lies well above 10. Indications are that uplift plays a central role.  ALMA has shown that molecular clouds are moving well below free-fall speeds.  Motivated by these results, we suggested in \cite{McNamara2016} that lifting parcels of gas that then fall inward in less than the free-fall speed can promote thermally unstable cooling, pushing $\text{min}(\text{t}_{\text{cool}}/\text{t}_{\text{I}})$ toward unity. Infall timescales may be significantly larger than the free-fall timescale, with the limiting timescale governed by the terminal speed of the clouds.  This conjecture, which we have dubbed ``stimulated feedback'', posits that the AGN itself stimulates the cooling that fuels it. 
Similarly,  \citet{Pizzolato2005} suggested, similarly, that inhomogeneities created by jets and bubbles would initiate non-linear cooling  eventually leading to molecular gas condensation. \citet{Voit2016} have added an uplift mechanism to their precipitation model.  While uplift has emerged as a significant element of thermally unstable cooling, the relevance of the  minimum value of $\text{t}_{\text{cool}}/\text{t}_{\text{ff}}$ to thermally unstable cooling is, in our view, less clear.  This conjecture is difficult to test because the gravitational potential wells of elliptical and brightest cluster galaxies are so similar that the free-fall timescales are nearly identical from system to system \citep{Hogan2017}.  Therefore, most studies have exagerated the correlation between $\text{t}_{\text{cool}}/\text{t}_{\text{ff}}$ and thermally unstable cooling which is instead driven by the cooling time.  We cannot exclude a central role for the
free-fall time.  But observation provides little indication that a specific value or range of this ratio is driving factor.   

Feedback stimulated by uplift may be a general phenomenon. Cooling may be stimulated by any mechanism that lifts low entropy gas out from the core of a galaxy. Mechanisms include rising radio bubbles, jets, or atmospheric sloshing initiated by a merger.

\begin{figure}[t!]
\includegraphics[width=0.49\textwidth]{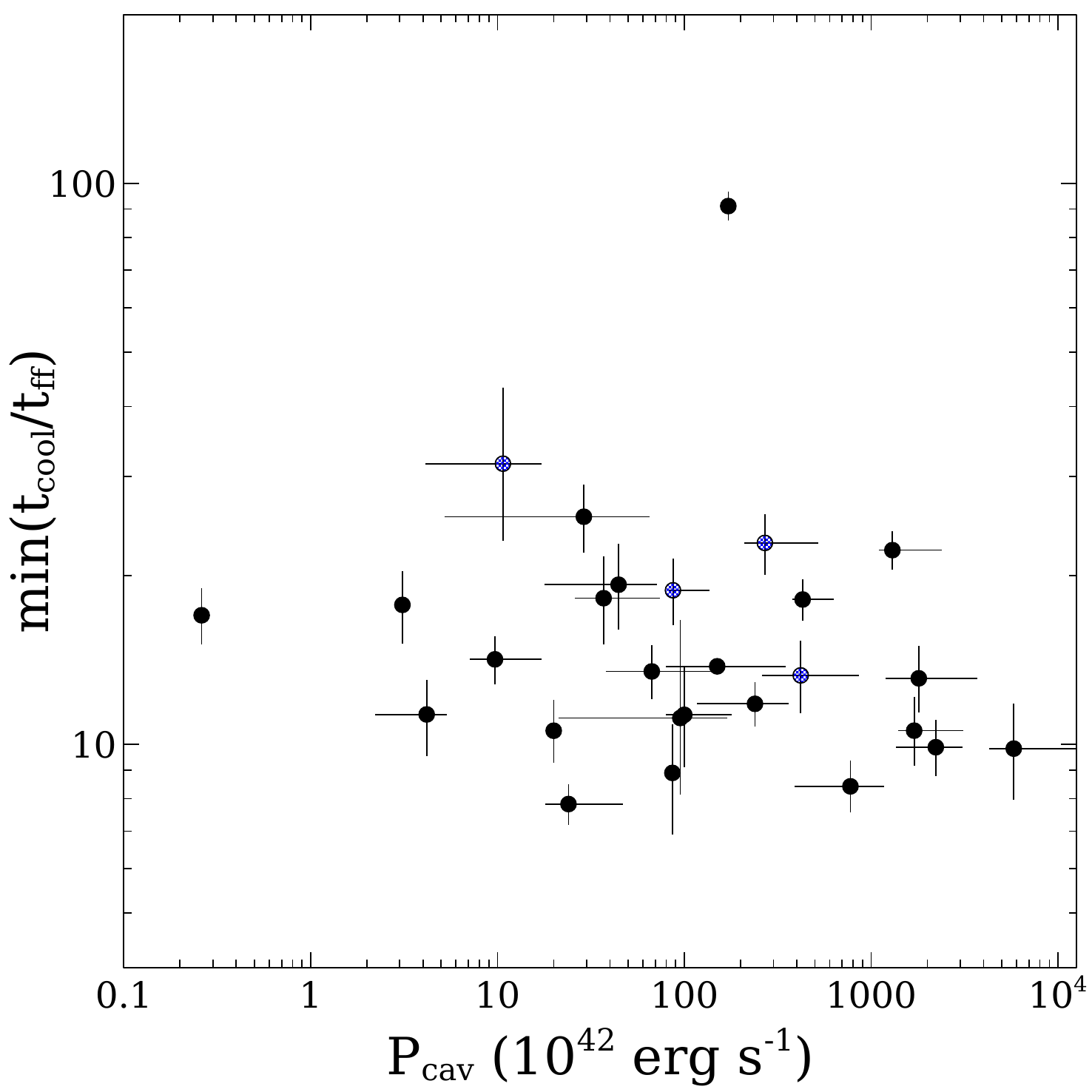}
\caption{Minimum cooling to free-fall ratio vs. cavity power. Black symbols denote systems observed
with CO emission while blue symbols denote upper limits. \label{fig:ratio_vs_pcav}}
\end{figure}

\section{Summary} \label{sec:conclusion}

We investigated the molecular cloud properties of 55 central cluster galaxies with molecular gas masses lying between $10^{8}-10^{11}\ \text{M}_{\odot}$. 
Chandra X-ray observations were used to measure cluster mass profiles into the central 10 kpc of the clusters accounting for the mass of the central galaxy.  Acceleration profiles, temperature, density, and other thermodynamic parameters were used to examine the possibility that the molecular gas formed through thermally unstable cooling from the hot atmosphere. 

\begin{itemize}

\item Molecular gas at levels between $10^{9}$ $\rm M_\odot$ and $10^{11}~\rm M_\odot$ is preferentially observed in BCGs with cooling times less than 1 Gyr, or entropy index below 35 cm$^{2}$ at a resolved radius of 10 kpc.  The corresponding star formation rates lie between 0.55 and 270 $\text{M}_{\odot}\ \text{yr}^{-1}$. 

\item Molecular gas mass is strongly and positively correlated with both central atmospheric gas density and atmospheric gas mass. These trends are consistent with the molecular clouds having condensed from the hot atmospheres.

\item The molecular gas depletion timescales due to star formation lie below1 Gyr.  The average depletion timescale $\simeq 5\times 10^8~\rm yr$ is a constant independent of molecular gas mass.  This long timescale is in tension with ALMA observations showing the molecular gas has formed recently and has not relaxed to a dynamically stable configuration such as a disk or ring.

\item Central galaxies rich in molecular gas lie in the range $10 \lesssim \text{min}(\text{t}_{\text{cool}}/\text{t}_{\text{ff}}) \lesssim 25$.  This small observed range can be plausibly attributed to an observational selection effect, although real physical effects cannot be excluded.  The range is inconsistent with models positing that thermally unstable cooling ensues when $\text{min}(\text{t}_{\text{cool}}/\text{t}_{\text{ff}}) \lesssim 10$, but is broadly consistent with the range found in chaotic cold accretion models \citep{Gaspari2012}.
 
\item Large fluctuations in the central atmospheric gas density and temperature in response to AGN activity found in many feedback models are absent. Accordingly, no correlation between $\text{min}(\text{t}_{\text{cool}}/\text{t}_{\text{ff}})$ and molecular gas mass is found.  

\item The tendency for molecular clouds to lie behind buoyantly rising X-ray bubbles suggests the molecular gas is being lifted directly and/or is condensing from thermally-unstable gas being lifted by the bubbles. We suggested here and elsewhere that the feedback loop may be stimulated by uplift from the X-ray bubbles themselves, and in some instances by ram pressure induced by atmospheric sloshing.

\end{itemize}

\vspace{2cm}
\acknowledgments

BRM, FAP, ACE, MTH, ANV, HRR, PEJN, IB, and PS acknowledge funding from the Natural Science and Engineering Research Council of Canada and from the Canadian Space Agency. ACE acknowledges support from STFC grant ST/P00541/1. HRR acknowledges support from ERC Advanced Grant Feedback 340442. Support for this work was provided in part by the National Aeronautics and Space Administration through Chandra Award Number G05-16134X issued by the Chandra X-ray Observatory Center, which is operated by the Smithsonian Astrophysical Observatory for and on behalf of the National Aeronautics Space Administration under contract NAS8-03060 We also acknowledge IRAM observers Pierre Hily-Blant, Richard Wilman, Stephen Hamer and Raymond Oonk. The scientific results reported in this article are based on observations made by the Chandra X-ray Observatory (CXO), Institut de Radioastronomie Millim\'{e}trique (IRAM) 30m telescope, and Two Micron All Sky Survey (2MASS). The following software from the Chandra X-ray Center (CXC) has been used: CIAO, ChIPs, and Sherpa. This research has also made use of the NASA/IPAC Extragalactic Database (NED). The plots in this paper were created using Veusz by Jeremy Sanders. 

\bibliographystyle{apj}
\bibliography{paper} 

\appendix

\section{Tables}

\begin{deluxetable*}{lcccccc}
  \tabletypesize{\normalsize}
  \tablecaption{Data Sample \label{tab:sample_bcg_coordinates}}
  \tablehead{
    \colhead{}      &   & \multicolumn2c{X-R\textsc{ay} C\textsc{ore} (J2000.0)} & \phantom{abc} & \multicolumn2c{BCG C\textsc{ore} (J2000.0)}  \\
    \colhead{Cluster} &  z  & \colhead{$\alpha$} & \colhead{$\delta$} & BCG N\textsc{ame} & \colhead{$\alpha$} & \colhead{$\delta$}
  }
  \startdata
    A85\dotfill\dotfill         & 0.055     &   00:41:50.567  & -9:18:10.86     &   MCG-02-02-086             & 00:41:50.524  & -09:18:10.94      \\
    A262\dotfill                & 0.017     &   01:52:46.299  & +36:09:11.80    &   NGC708                    & 01:52:46.482  & +36:09:06.53    \\
    A478\dotfill                & 0.088     &   04:13:25.345  & +10:27:55.15    & 2MASXJ04132526+1027551      & 04:13:25.266  & +10:27:55.14  \\
    A496\dotfill                & 0.033     &   04:33:38.038  & -13:15:39.65    & MCG-02-12-039               & 04:33:37.841  & -13:15:43.04  \\
    A1060\dotfill               & 0.013     &   10:36:42.830  & -27:31:39.62    & NGC3311                     & 10:36:42.821  & -27:31:42.02  \\
    A1068\dotfill               & 0.138     &   10:40:44.520  & +39:57:10.28    & 2MASXJ10404446+3957117      & 10:40:44.504  & +39:57:11.26  \\
    A1664\dotfill               & 0.128     &   13:03:42.622  & -24:14:41.59    &   2MASXJ13034252-2414428    & 13:03:42.521  & -24:14:42.81  \\
    A1835\dotfill               & 0.253     &   14:01:01.951  & +02:52:43.18    & 2MASXJ14010204+0252423      & 14:01:02.043  & +02:52:42.34  \\
    A1991\dotfill               & 0.059     &   14:54:31.553  & +18:38:39.79    & NGC5778                     & 14:54:31.465  & +18:38:32.57  \\
    A2052\dotfill               & 0.035     &   15:16:44.443  & +7:01:17.32     & UGC9799                     & 15:16:44.487  & +07:01:18.00  \\
    A2204\dotfill               & 0.152     &   16:32:46.920  & +05:34:32.86    & VLSSJ1632.7+0534            & 16:32:46.94   & +05:34:32.6   \\
    A2597\dotfill               & 0.085     &   23:25:19.779  & -12:07:27.63    & PKS2322-12                  & 23:25:19.731  & -12:07:27.51  \\
    A3581\dotfill               & 0.023     &   14:07:29.791  & -27:01:04.06    & IC4374                      & 14:07:29.780  & -27:01:04.39    \\
    A3880\dotfill               & 0.058     &   22:27:54.455  & -30:34:32.88    & PKS2225-308                 & 22:27:54.463  & -30:34:32.12  \\
    Cygnus-A\dotfill            & 0.056     &   19:59:28.259  & +40:44:02.10    & CygnusA                     & 19:59:28.357  & +40:44:02.10  \\
    H1821+643\dotfill           & 0.297     &   18:21:57.191  & +64:20:36.56    & H1821+643                   & 18:21:57.237  & +64:20:36.23  \\
    Hydra-A\dotfill             & 0.055     &   9:18:05.673   & -12:05:43.65    & Hydra-A                     & 09:18:05.651  & -12:05:43.99  \\
    MACS1532.9+3021\dotfill     & 0.345     &   15:32:53.820  & +30:20:59.75    & SDSSJ153253.78+302059.3     & 15:32:53.778  & +30:20:59.42  \\
    NGC4325\dotfill             & 0.026     &   12:23:06.659  & +10:37:15.53    & NGC4325                     & 12:23:06.672  & +10:37:17.05  \\
    NGC5044\dotfill             & 0.009     &   13:15:23.904  & -16:23:07.53    & NGC5044                     & 13:15:23.969  & -16:23:08.00  \\
    PKS0745-191\dotfill         & 0.103     &   7:47:31.228   & -19:17:41.01    & PKS0745-191                 & 07:47:31.296  & -19:17:40.34  \\
    RXCJ0338.6+0958\dotfill     & 0.036     &   3:38:41.055   & +9:58:02.26     &   2MASXJ03384056+0958119    & 3:38:40.579   & +9:58:11.78   \\
    RXCJ0352.9+1941\dotfill     & 0.109     &   3:52:59.001   & +19:40:59.81    & 2MASXJ03525901+1940595      & 3:52:59.016   & +19:40:59.59  \\
    RXJ0821.0+0752\dotfill      & 0.110     &   8:21:02.018   & +7:51:47.58     & 2MASXJ08210226+0751479      & 08:21:02.265  & +07:51:47.95  \\
    RXJ1504.1-0248\dotfill      & 0.215     &   15:04:07.529  & -2:48:16.75     & 2MASXJ15040752-0248161      & 15:04:07.519  & -02:48:16.65  \\
    RXCJ1524.2-3154\dotfill     & 0.103     &   15:24:12.861  & -31:54:23.52    & 2MASXJ15241295-3154224      & 15:24:12.957  & -31:54:22.45  \\
    RXCJ1558.3-1410\dotfill     & 0.097     &   15:58:21.948  & -14:09:58.43    & PKS1555-140                 & 15:58:21.948  & -14:09:59.05  \\
    RXJ1350.3+0940\dotfill      & 0.090     &   13:50:21.891  & +9:40:10.84     & 2MASXJ13502209+0940109      & 13:50:22.136  & +09:40:10.66  \\
    RXCJ1459.4-1811\dotfill     & 0.236     &   14:59:28.713  & -18:10:45.01    & 2MASXJ14592875-1810453      & 14:59:28.763  & -18:10:45.19  \\
    ZwCl1883\dotfill            & 0.194     &   8:42:55.952   & +29:27:25.61    & 2MASXJ08425596+2927272      & 08:42:55.972  & +29:27:26.91  \\
    ZwCl3146\dotfill            & 0.291     &   10:23:39.741  & +4:11:10.64     & 2MASXJ10233960+0411116      & 10:23:39.609  & +04:11:11.68  \\
    ZwCl7160\dotfill            & 0.258     &   14:57:15.073  & +22:20:35.18    & 2MASXJ14571507+2220341      & 14:57:15.077  & +22:20:34.16  \\
    ZwCl8276\dotfill            & 0.076     &   17:44:14.448  & +32:59:29.38    & 2MASXJ17441450+3259292      & 17:44:14.5  &  +32:59:29    \\
    4C+55.16\dotfill            & 0.242     &   8:34:54.917   & +55:34:21.44    & 2MFGC06756                  & 08:34:54.903  & +55:34:21.09  \\
    A1668\dotfill               & 0.063     &   13:03:46.602  & 13:03:46.602    & IC4130                      & 13:03:46.586  & +19:16:17.06  \\
    A2029\dotfill               & 0.077     &   15:10:56.104  & +5:44:41.14     & IC1101                      & 15:10:56.104  & +05:44:41.69  \\
    A2142\dotfill               & 0.091     &   15:58:20.880  & +27:13:44.21    & 2MASXJ15582002+2714000      & 15:58:20.028  & +27:14:00.06  \\
    A2151\dotfill               & 0.037     &   16:04:35.758  & +17:43:18.54    & NGC6041                     & 16:04:35.757  & +17:43:17.20  \\
    A2199\dotfill               & 0.030     &   16:28:38.249  & +39:33:04.28    & NGC6166                     & 16:28:38.276  & +39:33:04.97  \\
    A2261\dotfill               & 0.224     &   17:22:27.140  & +32:07:57.43    & 2MASXJ17222717+3207571      & 17:22:27.173  & +32:07:57.18  \\
    A2319\dotfill               & 0.056     &   19:21:09.638  & +43:57:21.53    & MCG+07-40-004               & 19:21:10.049  & +43:56:44.32  \\
    A2390\dotfill               & 0.228     &   21:53:36.768  & +17:41:42.17    & 2MASXJ21533687+1741439      & 21:53:36.827  & +17:41:43.73  \\
    A2462\dotfill               & 0.073     &   22:39:11.367  & -17:20:28.33    & 2MASXJ22391136-1720284      & 22:39:11.367  & -17:20:28.49  \\
    A2634\dotfill               & 0.031     &   23:38:29.426  & +27:01:53.86    & NGC7720                     & 23:38:29.390  & +27:01:53.53  \\
    A2657\dotfill               & 0.040     &   23:44:57.253  & +09:11:30.74    & 2MASXJ23445742+0911349      & 23:44:57.422  & +09:11:34.96  \\
    A2626\dotfill               & 0.055     &   23:36:30.375  & +21:08:48.21    & IC5338                      & 23:36:30.482  & +21:08:47.46  \\
    A2665\dotfill               & 0.056     &   23:50:50.557  & +6:09:03.00     & MCG+01-60-039               & 23:50:50.537  & +06:08:58.35  \\
    A2734\dotfill               & 0.063     &   0:11:21.665   & -28:51:15.05    &   ESO409-25                 & 00:11:21.667  &   -28:51:15.85  \\
    A3526\dotfill               & 0.011     &   12:48:48.949  & -41:18:43.92    & NGC4696                     & 12:48:49.277  & -41:18:39.92  \\
    AWM7\dotfill                & 0.017     &   12:30:49.361  & +12:23:28.10    & NGC1129                     & 02:54:27.400  & +41:34:46.70  \\
    M87\dotfill                 & 0.004     &   12:30:49.368  & +12:23:28.50    & M87                         & 12:30:49.423  & +12:23:28.04  \\
    RXJ0439.0+0520\dotfill      & 0.208     &   4:39:02.180   & +5:20:43.33     & 2MASXJ04390223+0520443      & 04:39:02.263  & +05:20:43.70  \\
    RXJ1347.5-1145\dotfill      & 0.451     &   13:47:30.641  & -11:45:08.51    & GALEXJ134730.7-114509       & 13:47:31.00   & -11:45:09.0   \\
    ZwCl235\dotfill             & 0.083     &   0:43:52.184   & +24:24:20.09    & 2MASXJ00435213+2424213      & 00:43:52.140  & +24:24:21.31    \\
    ZwCl2089\dotfill            & 0.230     &   9:00:36.887   & +20:53:40.79    & 2MASXJ09003684+2053402      & 09:00:36.848  & +20:53:40.24  \\
  \enddata
  \tablecomments{BCG Coordinates were taken from \cite{Hogan2015}}
  
\end{deluxetable*}
\phantom{abc}

\begin{deluxetable*}{llccc}
    \tabletypesize{\tiny}
    \fontsize{7}{7}
    \selectfont
    \tablecaption{X-ray Observation Properties \label{tab:xray_observation_properties}}
    \tablehead{
        \colhead{}        &                  & \multicolumn2c{Total Exposure (ks)} \\
        \colhead{Cluster} & \colhead{ObsID}  & \colhead{Raw} & \colhead{Clean} 
    }
    \startdata
      A85\dotfill                  & 15173,15174,16263,16264,904               & 195.2   & 193.6 \\
      A262\dotfill                 & 2215,7921                                 & 139.4   & 137.4 \\
      A478\dotfill                 & 1669,6102                                 & 52.4    & 46.8 \\
      A496\dotfill                 & 4976,931                                  & 94.0    & 61.7 \\
      A1060\dotfill                & 2220                                      & 31.9    & 29.4  \\
      A1068\dotfill                & 1652                                      & 26.8    & 23.2 \\
      A1664\dotfill                & 1648,17172,17173,17557,17568,7901         & 245.5   & 233.3 \\
      A1835\dotfill                & 6880,6881,7370                            & 193.7   & 139.1 \\
      A1991\dotfill                & 3193                                      & 38.3    & 34.5 \\
      A2052\dotfill                & 10477,10478,10479,10480,10879,10914,10915,10916,10917,5807,890 & 654.0 & 640.4 \\
      A2204\dotfill                & 6104,7940                                 & 86.8    & 80.1 \\
      A2597\dotfill                & 6934,7329,922                             & 151.6   & 137.6 \\
      A3581\dotfill                & 12884,1650                                & 91.7    & 90.6 \\
      A3880\dotfill                & 5798                                      & 22.3    & 18.6 \\
      Cygnus-A\dotfill             & 5830,5831,6225,6226,6228,6229,6250,6252   & 198.1   & 193.6 \\
      H1821+643\dotfill            & 9398,9845,9846,9848                       & 87.0    & 83.2 \\
      Hydra-A\dotfill              & 4969,4970,576                             & 215.3   & 186.4 \\
      MACS1532.9+3021\dotfill      & 14009,1649,1665                           & 108.2   & 102.4 \\
      NGC4325\dotfill              & 3232                                      & 30.1    & 25.7 \\
      NGC5044\dotfill              & 9399                                      & 82.7    & 82.5 \\
      PKS0745-191\dotfill          & 12881,1509,2427,510                       & 220.6   & 210.1 \\
      RXCJ0338.6+0958\dotfill      & 7939,9792                                 & 83.3    & 81.2 \\
      RXCJ0352.9+1941\dotfill      & 10466                                     & 27.2    & 27.2 \\
      RXJ0821.0+0752\dotfill       & 17194,17563                               & 66.6    & 63.5 \\
      RXJ1504.1-0248\dotfill       & 17197,17669,17670,4935,5793               & 161.7   & 135.3 \\
      RXCJ1524.2-3154\dotfill      & 9401                                      & 40.9    & 40.9 \\
      RXCJ1558.3-1410\dotfill      & 9402                                      & 40.1    & 35.8 \\
      RXJ1350.3+0940\dotfill       & 14021                                     & 19.8    & 19.4 \\
      RXCJ1459.4-1811\dotfill      & 9428                                      & 39.6    & 39.5 \\
      ZwCl1883\dotfill             & 2224                                      & 29.8    & 26.3 \\
      ZwCl3146\dotfill             & 1651,9371                                 & 206.0   & 189.6 \\
      ZwCl7160\dotfill             & 4192,543                                  & 101.7   & 80.0 \\
      ZwCl8276\dotfill             & 11708,8267                                & 53.5    & 53.2 \\
      4C+55.16\dotfill             & 4940                                      & 96.0    & 65.5 \\
      A1668\dotfill                & 12877                                     & 10.0    & 10.0 \\
      A2029\dotfill                & 4977,6101,891                             & 107.6   & 103.3 \\
      A2142\dotfill                & 15186,16564,16565,5005                    & 199.7   & 184.6 \\
      A2151\dotfill                & 4996                                      & 21.8    & 14.4 \\
      A2199\dotfill                & 10748,10803,10804,10805,497,498           & 158.2   & 155.8 \\
      A2261\dotfill                & 5007                                      & 24.3    & 22.1 \\
      A2319\dotfill                & 15187,3231                                & 89.6    & 86.8 \\
      A2390\dotfill                & 4193,500,501                              & 113.9   & 88.2 \\
      A2462\dotfill                & 4159                                      & 39.2    & 37.6 \\
      A2626\dotfill                & 16136,3192                                & 135.6   & 132.5 \\
      A2634\dotfill                & 4816                                      & 49.5    & 47.5 \\
      A2657\dotfill                & 4941                                      & 16.1    & 15.9 \\
      A2665\dotfill                & 12280                                     & 9.9     & 9.4 \\
      A2734\dotfill                & 5797                                      & 19.9    & 18.9 \\
      A3526\dotfill                & 16223,16224,16225,16534,16607,16608,16609,16610 & 486.3  & 478.5 \\
      AWM7\dotfill                 & 11717,12016,12017,12018                   & 133.8   & 133.5 \\
      M87\dotfill                  & 5826,5827                                 & 283.0   & 283.0 \\
      RXJ0439.0+0520\dotfill       & 9369,9761                                 & 28.5    & 25.9 \\
      RXJ1347.5-1145\dotfill       & 13516,13999,14407,2222,3592,506,507       & 326.5   & 286.4 \\
      ZwCl2089\dotfill             & 10463,7897                                & 49.7    & 46.9 \\
      ZwCl235\dotfill              & 11735                                     & 19.8    & 19.4 \\
    \enddata
    \tablecomments{The hydrogen column density was frozen to these values (taken from \citealt{Kalberla2005}) when fitting in \textsc{xspec} with the \textsc{mekal} model unless the best fit value was found to be significantly different (these are marked with the asterisk (*) symbol).
}
\end{deluxetable*}
\phantom{abc}

\begin{deluxetable*}{lcccc}
  \tabletypesize{\normalsize}
	\tablecaption{Mass Parameters \label{tab:mass_properties}}
	\tablehead{
		\colhead{Cluster} &  \colhead{$\sigma_{*}$}			& \colhead{$r_{s}$} 		 	& \colhead{$A = 4\pi G\rho_{0}r_{s}2\mu m_{H}$}  	& \colhead{$M_{2500}$}   \\
						  &  \colhead{(km s$^{-1}$)}		& \colhead{(kpc)}  			 	& \colhead{(keV)}  		 	& \colhead{($10^{13}$ M$_{\odot}$)} 
	}
	\startdata
  A85\dotfill         &   $270\pm6$ & $376.2^{+37.0}_{-25.4}$   &   $49.1^{+4.0}_{-2.7}$    &   $22.2^{+1.1}_{-1.2}$ \\
  A262\dotfill        &   $189\pm3$ & $185.8^{+3.7}_{-0.4}$   &   $12.9^{+0.3}_{-0.0}$    &   $3.4^{+0.1}_{-0.1}$ \\
  A478\dotfill        &   $271\pm7$ & $588.2^{+206.8}_{-130.7}$ &   $71.4^{+18.2}_{-12.4}$    &   $33.3^{+5.4}_{-6.0}$ \\
  A496\dotfill        &   $228\pm5$ & $190.1^{+68.1}_{-38.3}$   &   $32.8^{+2.5}_{-1.5}$    &   $12.9^{+1.1}_{-1.1}$ \\
  A1060\dotfill       &   $208\pm12$  & $191.7$                 &   $36.9$                  &   $15.0$ \\
  A1068\dotfill       &   $311\pm12$  & $519.4^{+122.2}_{-79.8}$  &   $47.3^{+8.4}_{-5.0}$    &   $18.5^{+1.8}_{-1.7}$ \\
  A1664\dotfill       &   $267\pm12$  & $300.2^{+31.4}_{-29.5}$   &   $25.7^{+2.3}_{-1.1}$    &   $8.8^{+0.5}_{-0.5}$ \\
  A1835\dotfill       &   $486\pm24$  & $550.3^{+45.3}_{-61.8}$   &   $94.3^{+5.7}_{-7.7}$    &   $55.8^{+3.3}_{-3.2}$ \\
  A1991\dotfill       &   $222\pm8$ & $266.4^{+93.5}_{-98.9}$   &   $25.1^{+5.0}_{-6.7}$    &   $8.5^{+1.5}_{-1.4}$ \\
  A2052\dotfill       &   $221\pm5$ & $170.6^{+55.9}_{-39.6}$   &   $24.8^{+4.1}_{-3.3}$    &   $8.7^{+1.4}_{-1.4}$ \\
  A2204\dotfill       &   $343\pm13$  & $409.6^{+40.6}_{-36.0}$   &   $80.9^{+5.0}_{-4.3}$    &   $45.7^{+2.2}_{-2.4}$ \\
  A2597\dotfill       &   $218\pm10$  & $257.4^{+52.5}_{-16.9}$   &   $37.9^{+5.5}_{-2.0}$    &   $15.1^{+1.6}_{-1.6}$ \\
  A3581\dotfill       &   $195\pm3$ & $80.7^{+14.5}_{-13.6}$    &   $8.4^{+0.5}_{-0.5}$     &   $2.0^{+0.1}_{-0.1}$ \\
  A3880\dotfill       &   $236\pm7$ & $122.1^{+95.3}_{-47.4}$   &   $21.7^{+4.9}_{-2.8}$    &   $7.2^{+1.6}_{-1.6}$ \\
  Cygnus-A\dotfill      &   $268\pm8$ & $145.0^{+69.9}_{-43.1}$   &   $45.2^{+4.5}_{-2.1}$    &   $19.6^{+0.1}_{-0.1}$ \\
  H1821+643\dotfill     &   $250\pm15$  & $171.5^{+216.7}_{-15.9}$  &   $23.1^{+13.6}_{-3.7}$   &   $7.2^{+2.8}_{-2.7}$ \\
  Hydra-A\dotfill       &   $237\pm8$ & $551.8^{+22.7}_{-36.7}$   &   $37.8^{+0.9}_{-1.4}$    &   $12.2^{+0.2}_{-0.2}$ \\
  MACS1532.9+3021\dotfill   &   $250\pm15$  & $769.0^{+535.9}_{-144.0}$ &   $105.2^{+50.0}_{-14.8}$   &   $43.7^{+7.3}_{-7.5}$ \\
  NGC4325\dotfill       &   $174\pm5$ & $66.2^{+12.4}_{-7.6}$   &   $4.9^{+0.5}_{-0.3}$     &   $1.0^{+0.1}_{-0.1}$ \\
  NGC5044\dotfill       &   $196\pm11$  & $45.1^{+5.4}_{-4.7}$    &   $7.0^{+0.2}_{-0.2}$     &   $1.5^{+0.1}_{-0.1}$ \\
  PKS0745-191\dotfill     &   $290\pm14$  & $437.9^{+186.1}_{-115.8}$ &   $67.4^{+18.7}_{-12.5}$    &   $33.9^{+5.8}_{-5.5}$ \\
  RXCJ0338.6+0958\dotfill   &   $220\pm5$ & $153.2^{+62.5}_{-59.2}$   &   $21.3^{+4.6}_{-3.8}$    &   $7.1^{+1.6}_{-1.5}$ \\
  RXCJ0352.9+1941\dotfill   &   $239\pm10$  & $223.3^{+37.6}_{-15.0}$   &   $22.6^{+2.0}_{-0.9}$    &   $7.5^{+0.4}_{-0.4}$ \\
  RXJ0821.0+0752\dotfill    &   $247\pm9$ & $268.8^{+453.5}_{-159.1}$ &   $20.7^{+22.2}_{-7.9}$   &   $6.5^{+2.8}_{-2.6}$ \\
  RXJ1504.1-0248\dotfill    &   $386\pm22$  & $787.8^{+142.3}_{-107.2}$ &   $111.4^{+15.3}_{-12.0}$   &   $58.6^{+4.6}_{-4.5}$ \\
  RXCJ1524.2-3154\dotfill   &   $265\pm12$  & $450.5^{+104.3}_{-53.7}$  &   $64.5^{+11.1}_{-5.3}$   &   $30.9^{+3.0}_{-2.7}$ \\
  RXCJ1558.3-1410\dotfill   &   $280\pm14$  & $451.5^{+86.3}_{-89.4}$   &   $47.5^{+6.3}_{-6.4}$    &   $19.6^{+1.8}_{-1.8}$ \\
  RXJ1350.3+0940\dotfill    &   $188\pm13$  & $111.4^{+22.1}_{-59.2}$   &   $29.3^{+2.2}_{-9.3}$    &   $9.7^{+1.7}_{-1.8}$ \\
  RXCJ1459.4-1811\dotfill   &   $439\pm22$  & $421.1^{+128.7}_{-82.1}$  &   $46.1^{+8.3}_{-6.0}$    &   $21.6^{+2.1}_{-2.3}$ \\
  ZwCl1883\dotfill      &   $335\pm12$  & $315.6^{+196.8}_{-82.1}$  &   $27.7^{+9.6}_{-4.8}$    &   $10.3^{+1.4}_{-1.5}$ \\
  ZwCl3146\dotfill      &   $372\pm33$  & $719.6^{+104.5}_{-121.6}$ &   $87.1^{+10.0}_{-11.0}$    &   $38.2^{+2.9}_{-2.7}$ \\
  ZwCl7160\dotfill      &   $428\pm21$  & $455.3^{+102.0}_{-64.2}$  &   $67.3^{+10.4}_{-7.4}$   &   $34.4^{+3.3}_{-3.3}$ \\
  ZwCl8276\dotfill      &   $219\pm7$ & $531.6^{+59.4}_{-58.5}$   &   $54.3^{+4.8}_{-4.6}$    &   $21.5^{+1.2}_{-1.3}$ \\
  4C+55.16\dotfill      &   $274\pm24$  & $452.5^{+35.4}_{-27.3}$   &   $49.4^{+1.8}_{-1.8}$    &   $18.6^{+1.2}_{-1.1}$ \\
  A1668\dotfill       &   $226\pm7$ & $93.7^{+149.9}_{-14.4}$   &   $13.7^{+15.1}_{-1.2}$   &   $3.9^{+0.8}_{-0.7}$ \\
  A2029\dotfill       &   $336\pm10$  & $511.4^{+50.4}_{-30.9}$   &   $79.8^{+5.1}_{-3.1}$    &   $44.8^{+1.8}_{-1.9}$ \\
  A2142\dotfill       &   $241\pm11$  & $345.5^{+37.4}_{-21.3}$   &   $46.8^{+3.0}_{-1.5}$    &   $20.0^{+0.9}_{-0.8}$ \\
  A2151\dotfill       &   $219\pm4$ & $196.0^{+38.7}_{-55.3}$   &   $15.4^{+1.4}_{-2.3}$    &   $4.6^{+0.4}_{-0.4}$ \\
  A2199\dotfill       &   $246\pm4$ & $364.3^{+181.3}_{-119.4}$ &   $48.3^{+16.7}_{-11.0}$    &   $21.6^{+5.3}_{-5.6}$ \\
  A2261\dotfill       &   $460\pm17$  & $396.6^{+114.4}_{-98.9}$  &   $77.0^{+13.1}_{-7.4}$   &   $45.1^{+5.2}_{-4.8}$ \\
  A2319\dotfill       &   $249\pm7$ & $397.8^{+110.7}_{-316.9}$ &   $54.6^{+11.9}_{-41.6}$    &   $25.2^{+5.0}_{-4.6}$ \\
  A2390\dotfill       &   $348\pm22$  & $799.1^{+183.0}_{-79.7}$  &   $101.3^{+17.8}_{-7.4}$    &   $47.6^{+3.9}_{-4.0}$ \\
  A2462\dotfill       &   $260\pm8$ & $458.7^{+368.9}_{-207.7}$ &   $28.7^{+14.7}_{-8.5}$   &   $8.8^{+1.8}_{-1.8}$ \\
  A2634\dotfill       &   $269\pm3$ & $133.9^{+63.2}_{-62.7}$   &   $38.4^{+10.3}_{-13.7}$    &   $15.9^{+5.5}_{-5.9}$ \\
  A2626\dotfill       &   $243\pm7$ & $248.9^{+25.5}_{-25.6}$   &   $22.1^{+1.3}_{-1.3}$    &   $7.4^{+0.3}_{-0.3}$ \\
  A2657\dotfill       &   $172\pm6$ & $103.8^{+98.6}_{-64.1}$   &   $8.8^{+2.0}_{-2.8}$     &   $2.1^{+0.6}_{-0.6}$ \\
  A2665\dotfill       &   $248\pm7$ & $613.7^{+464.7}_{-203.9}$ &   $35.1^{+19.3}_{-8.7}$   &   $10.0^{+0.0}_{-0.0}$ \\
  A2734\dotfill       &   $231\pm8$ & $379.3^{+827.8}_{-110.3}$ &   $21.1^{+27.0}_{-4.0}$   &   $5.8^{+1.2}_{-1.2}$ \\
  RXJ0439.0+0520\dotfill    &   $389\pm21$  & $706.1^{+387.0}_{-237.1}$ &   $57.7^{+25.8}_{-16.0}$    &   $22.1^{+4.6}_{-4.3}$ \\
  RXJ1347.5-1145\dotfill    &   $250\pm15$  & $308.9^{+40.9}_{-31.6}$   &   $153.5^{+11.7}_{-10.5}$   &   $94.7^{+7.1}_{-7.1}$ \\
  ZwCl235\dotfill       &   $240\pm8$ & $206.1^{+63.8}_{-39.3}$   &   $21.4^{+2.3}_{-1.8}$    &   $7.2^{+0.6}_{-0.5}$ \\
  ZwCl2089\dotfill      &   $296\pm18$  & $245.4^{+25.6}_{-28.3}$   &   $31.6^{+1.4}_{-2.0}$    &   $11.9^{+0.5}_{-0.6}$ \\
	\enddata
	\tablenotetext{a}{$\sigma_{*}$ denote the equivalent velocity dispersion of the central galaxy inferred from 2MASS isophotal magnitude if the galaxy consisted only of its stars.}
	\tablenotetext{b}{$r_{s}$ and $\rho_{0}$ denote the characteristic scale radius and density of the NFW profile obtained from the \textsc{isonfwmass} model (See Section \ref{sub:mass_profile}).}
  \tablenotetext{c}{$M_{2500}$ denote the total cluster mass}
\end{deluxetable*}
\phantom{abc}

\begin{deluxetable*}{lccccccl}
  \tabletypesize{\footnotesize}
	\tablecaption{Cavity Power and Star Formation Rate \label{tab:cavity_power_and_sfr}}
	\tablehead{
		\colhead{} 		  & & \multicolumn2c{Cavity Power} & &  \multicolumn2c{Star Formation}  &  \\
		\cmidrule{3-4} \cmidrule{6-7}
		\colhead{Cluster} & \colhead{$L_{radio}$}              & \colhead{$P_{mech}$}     & \colhead{$P_{cav}$}                &  \colhead{$L_{H\alpha}$}            & \colhead{$\text{SFR}_{\text{H}\alpha}$}   & \colhead{$\text{SFR}_{\text{IR}}$}       & \colhead{Ref.}  \\
                      & \colhead{($10^{37}$ erg s$^{-1}$)} & \colhead{($10^{42}$ erg s$^{-1}$)} & \colhead{($10^{42}$ erg s$^{-1}$)} &  \colhead{($10^{40}$ erg s$^{-1}$)} & \colhead{(M$_{\odot}\text{yr}^{-1}$)}     & \colhead{(M$_{\odot}\text{yr}^{-1}$)}    & 
	}
	\startdata
  A85\dotfill               &  $2550\pm110$             &  $35.9\pm1.7$       & $37.0_{-11.0}^{+37.0}$        & $0.43$    & $0.033\pm0.010$   & $1.57$    & [1],[2] \\
  A262\dotfill              &  $277\pm10$               &  $12.4\pm2.3$       & $9.7_{-2.6}^{+7.5}$           & $0.43$    & $0.033\pm0.010$   & $0.55$    & [1],[3] \\
  A478\dotfill              &  $4420\pm176$             &  $46.7\pm0.5$       & $100.0_{-20.0}^{+80.0}$       & $7.86$    & $1.453\pm0.186$   & -         & [1],[3] \\
  A496\dotfill              &  $1840\pm65$              &  $30.7\pm2.1$       & $0.261$                       & $0.72$    & $0.065\pm0.017$   & -         & [14],[2] \\
  A1060\dotfill             &  $1.7\pm1.1$              &  $1.1\pm0.5$        & $172.0$                       & -         & -                 & -         & [14],- \\
  A1068\dotfill             &  $6814\pm214$             &  $57.5\pm1.1$       & $20.0$                        & $121.43$  & $51.039\pm2.878$  & $187.45$  & [1],[3] \\
  A1664\dotfill             &  $9678\pm316$             &  $68.1\pm3.0$       & $95.2_{-74.0}^{+74.0}$        & $78.57$   & $28.982\pm1.862$  & $14.54$   & [4],[3] \\
  A1835\dotfill             &  $37100$                  &  $129.8\pm19.2$     & $1800.0_{-600.0}^{+1900.0}$   & $100.00$  & $39.654\pm2.370$  & -         & [1],[3] \\
  A1991\dotfill             &  $1973\pm65$              &  $31.7\pm2.1$       & $86.4$                        & $0.60$    & $0.051\pm0.014$   & $<1.66$   & [11],[2] \\
  A2052\dotfill             &  $97400\pm3700$           &  $206.3\pm47.0$     & $150.0_{-70.0}^{+200.0}$      & $1.38$    & $0.151\pm0.033$   & $1.37$    & [1],[2] \\
  A2204\dotfill             &  $26700\pm991$            &  $110.9\pm13.5$     & $775.0_{-385.0}^{+395.0}$     & $114.29$  & $47.171\pm2.709$  & $14.62$   & [5],[3] \\
  A2597\dotfill             &  $205300\pm6160$          &  $295.1\pm86.7$     & $67.0_{-29.0}^{+87.0}$        & $37.14$   & $10.943\pm0.880$  & -         & [1],[3] \\
  A3581\dotfill             &  $4720\pm166$             &  $48.2\pm0.3$       & $3.1$                         & $21.30$   & $5.311\pm0.505$   & -         & [9],[2] \\
  A3880\dotfill             &  $11300\pm342$            &  $73.4\pm4.1$       & $29.0_{-23.8}^{+36.2}$        & -         & -                 & -         & [5],- \\
  Cygnus-A\dotfill          &  $72800000\pm1870000$     &  $4939.9\pm4699.6$  & $1300.0_{-200.0}^{+1100.0}$   & $21.30$   & $5.311\pm0.505$   & -         & [1],[2] \\
  H1821+643\dotfill         &  $<5990$                  &  $<54.1\pm0.6$      & No Cavity                     & -         & -                 & -         & [12],- \\
  Hydra-A\dotfill           &  $1710000\pm54500$        &  $816.5\pm408.9$    & $430.0_{-50.0}^{+200.0}$      & $11.43$   & $2.364\pm0.271$   & -         & [1],[3] \\
  MACS1532.9+3021\dotfill   &  $56400\pm1940$           &  $158.7\pm28.9$     & $2220.0_{-860.0}^{+860.0}$    & $300.01$  & $165.407\pm7.110$ & $96.15$   & [10],[3] \\
  NGC4325\dotfill           &  $<32.1$                  &  $<4.4\pm1.3$       & -                             & $0.36$    & $0.026\pm0.008$   & $<0.66$   & -,[7] \\
  NGC5044\dotfill           &  $42.0\pm1.7$             &  $5.0\pm1.4$        & $4.2_{-2.0}^{+1.2}$           & -         & -                 & -         & [8],- \\
  PKS0745-191\dotfill       &  $387000\pm13700$         &  $400.2\pm141.1$    & $1700.0_{-300.0}^{+1400.0}$   & $140.00$  & $61.411\pm3.318$  & $17.07$   & [1],[6] \\
  RXCJ0338.6+0958\dotfill   &  $706\pm35$               &  $19.4\pm2.5$       & $24.0_{-6.0}^{+23.0}$         & $7.14$    & $1.283\pm0.169$   & $2.09$    & [1],[3] \\
  RXCJ0352.9+1941\dotfill   &  $<648$                   &  $<18.6\pm2.5$      & No Cavity                     & $41.43$   & $12.612\pm0.982$  & $11.04$   & [12],[3] \\
  RXJ0821.0+0752\dotfill    &  $728$                    &  $19.7\pm2.5$       & -                             & $30.00$   & $8.290\pm0.711$   & $36.91$   & -,[6] \\
  RXJ1504.1-0248\dotfill    &  $50800\pm2140$           &  $150.9\pm26.1$     & -                             & -         & -                 & -         & -,- \\
  RXCJ1524.2-3154\dotfill   &  $8180\pm261$             &  $62.8\pm2.0$       & $239.0_{-122.0}^{+122.0}$     & -         & -                 & -         & [5],- \\
  RXCJ1558.3-1410\dotfill   &  $66500\pm2350$           &  $171.8\pm33.6$     & $44.5_{-26.7}^{+26.7}$        & -         & -                 & -         & [5],- \\
  RXJ1350.3+0940\dotfill    &  $36000\pm1100$           &  $128.0\pm18.6$     & -                             & -         & -                 & -         & -,- \\
  RXCJ1459.4-1811\dotfill   &  $107000\pm3230$          &  $215.5\pm50.8$     & No Cavity                     & -         & -                 & -         & [12],- \\
  ZwCl1883\dotfill          &  $16400\pm587$            &  $87.8\pm7.4$       & -                             & -         & -                 & -         & -,- \\
  ZwCl3146\dotfill          &  $12000\pm814$            &  $75.6\pm4.6$       & $5800.0_{-1500.0}^{+6800.0}$  & $500.01$  & $321.332\pm11.850$& -       & [1],[3] \\
  ZwCl7160\dotfill          &  $19700\pm124$            &  $95.7\pm9.4$       & -                             & $35.72$   & $10.399\pm0.846$  & -         & -,[3] \\
  ZwCl8276\dotfill          &  $7850\pm282$             &  $61.6\pm1.8$       & -                             & $9.29$    & $1.805\pm0.220$   & $3.71$    & -,[3] \\
  4C+55.16\dotfill          &  $8870000\pm266000$       &  $1799.3\pm1230.9$  & $420.0_{-160.0}^{+440.0}$     & $71.43$   & $25.605\pm1.693$  & -         & [1],[3] \\
  A1668\dotfill             &  $4630\pm141$             &  $47.8\pm0.4$       & -                             & $12.00$   & $2.519\pm0.284$   & $<1.66$   & -,[6] \\
  A2029\dotfill             &  $48200\pm1850$           &  $147.2\pm24.9$     & $87.0_{-4.0}^{+49.0}$         & $0.80$    & $0.075\pm0.019$   & -         & [1],[6] \\
  A2142\dotfill             &  $<440$                   &  $<$$15.4\pm2.4$      & No Cavity                     & -         & -               & -         & [12],- \\
  A2151\dotfill             &  $207$                    &  $10.8\pm2.2$       & -                             & $5.80$    & $0.979\pm0.137$   & -         & -,[6] \\
  A2199\dotfill             &  $46700\pm1560$           &  $144.9\pm24.1$     & $270.0_{-60.0}^{+250.0}$      & $3.50$    & $0.508\pm0.083$   & -         & [1],[6] \\ 
  A2261\dotfill             &  $3060$                   &  $39.2\pm1.4$       & -                             & -         & -                 & -         & -,- \\
  A2319\dotfill             &  $<157$                   &  $<$$9.4\pm2.0$       & -                             & $10.00$   & $1.987\pm0.237$ & -         & -,[6] \\
  A2390\dotfill             &  $221000\pm7770$          &  $305.7\pm91.8$     & No Cavity                     & $44.29$   & $13.754\pm1.050$  & -         & [12],[7] \\
  A2462\dotfill             &  $<279$                   &  $<$$12.4\pm2.3$      & -                             & $5.80$    & $0.979\pm0.137$ & -         & -,[6] \\
  A2634\dotfill             &  $103000$                 &  $211.6\pm49.2$     & -                             & $3.70$    & $0.546\pm0.088$   & -         & -,[6] \\
  A2657\dotfill             &  $<80.1$                  &  $<$$6.8\pm1.7$       & -                             & $0.17$    & $0.010\pm0.004$ & -         & -,[2] \\
  A2626\dotfill             &  $2480\pm111$             &  $35.4\pm1.8$       & $10.7_{-6.6}^{+6.6}$          & $3.30$    & $0.470\pm0.078$   & $<1.66$   & [5],[6] \\
  A2665\dotfill             &  $2520\pm94$              &  $35.7\pm1.8$       & -                             & $0.60$    & $0.051\pm0.014$   & $<1.66$   & -,[6] \\
  A2734\dotfill             &  $651\pm34$               &  $18.6\pm2.5$       & -                             & -         & -                 & -         & -,- \\
  A3526\dotfill             &  $7010\pm194$             &  $58.3\pm1.3$       & No Cavity                     & $0.36$    & $0.026\pm0.008$   & -         & [13],[3] \\
  AWM7\dotfill              &  $<14.2$                  &  $<3.0\pm1.0$       & -                             & $0.36$    & $0.026\pm0.008$   & -         & -,[3] \\
  M87\dotfill               &  $34100\pm1210$           &  $124.6\pm17.6$     & -                             & $0.79$    & $0.073\pm0.019$   & -         & -,[3] \\
  RXJ0439.0+0520\dotfill    &  $85000\pm2990$           &  $193.3\pm41.8$     & -                             & $78.57$   & $28.982\pm1.862$  & $18.66$   & -,[3] \\
  RXJ1347.5-1145\dotfill    &  $217000\pm8650$          &  $302.9\pm90.5$     & No Cavity                     & $214.29$  & $106.805\pm5.079$ & -         & [12],[3] \\
  ZwCl235\dotfill           &  $5170\pm166$             &  $50.4\pm0.0$       & -                             & $2.93$    & $0.403\pm0.069$   & $<1.66$   & -,[7] \\
  ZwCl2089\dotfill          &  $8980\pm573$             &  $65.7\pm2.6$       & -                             & $71.43$   & $25.605\pm1.693$  & $270.47$  & -,[3] \\
	\enddata
  \tablecomments{\phantom{x} References for cavity power or H$\alpha$ luminosity -- [1] \cite{Rafferty2006}, [2] ACCEPT Database \cite{Cavagnolo2009}, [3] \cite{Edge2001}, [4] \cite{Kirkpatrick2009}, [5] Hlavacek-Larrondo (priv. comm, 2014), [6] \cite{Salome2003}, [7] \cite{Crawford1999}, [8] \cite{Cavagnolo2010}, [9] \cite{Canning2013}, [10] \cite{Hlavacek-Larrondo2013}, [11] \cite{Pandge2013}, [12] \cite{Shin2016}, [13] \cite{Panagoulia2014b}, [14] \cite{Birzan2012}}  
\end{deluxetable*}
\phantom{abc}


\end{document}